\documentstyle[11pt,aaspp4]{article}

\lefthead{Bak{\i}\c{s} et al.}
\righthead{Eclipsing Binaries towards Cyg~OB}

\begin{document}

\title{Study of Eclipsing Binary and Multiple Systems in OB Associations II. The Cygnus OB Region: V443~Cyg, V456~Cyg and V2107~Cyg}

\author{V.~Bak{\i}\c{s}}
\affil{Akdeniz University, Faculty of Science, Department of Space Sciences and
Technologies, Antalya, Turkey}

\author{H. Hensberge}
\affil{Royal Observatory of Belgium, Ringlaan 3, 1180 Brussels, Belgium}

\author{S. Bilir}
\affil{\.{I}stanbul University, Faculty of Science, Department of Astronomy and
Space Sciences, 34119 University-\.{I}stanbul, Turkey}

\author{H. Bak{\i}\c{s}}
\affil{Akdeniz University, Faculty of Science, Department of Space Sciences and
Technologies, Antalya, Turkey}

\author{F. Y{\i}lmaz}
\affil{\c{C}anakkale Onsekiz Mart University, Physics Department, Terzio\v{g}lu
Campus, TR-17020, \c{C}anakkale, Turkey}

\author{E. K{\i}ran}
\affil{Ege University, Faculty of Science, Department of Astronomy and Space
Sciences, \.Izmir, Turkey}

\author{O. Demircan}
\affil{\c{C}anakkale Onsekiz Mart University, Department of Space Sciences and
Technologies, Terzio\v{g}lu Campus, TR-17020, \c{C}anakkale, Turkey}

\author{M. Zejda}
\affil{Department of Theoretical Physics and Astrophysics, Masaryk University,
Brno, Czech Republic}

\and

\author{Z.~Mikul\'a\v sek}
\affil{Department of Theoretical Physics and Astrophysics, Masaryk University,
Brno, Czech Republic}

\begin{abstract}
Three presumably young eclipsing binary systems in the direction of the
Cygnus OB1, OB3 and OB9 associations are studied. Component spectra are
reconstructed and their orbits are determined using light curves and 
spectra disentangling techniques. V443\,Cyg and V456\,Cyg have 
circular orbits, while the light curve of V2107\,Cyg imposes a  
slightly eccentric orbit ($e=0.045\pm0.03)$. 
V443~Cyg harbours F--type stars, and not young early--A stars as 
previously suggested in the literature based on photometry solely. 
It appears to be situated in the foreground (distance $0.6\pm0.2$~kpc) 
of the young stellar populations in Cygnus. V456~Cyg, at a distance of 
$0.50\pm0.03$~kpc consists of a slightly metal-weak A--type and an  
early--F star. The age of both systems, on or very near to the main 
sequence, remains uncertain by an order of magnitude. V2107~Cyg is a 
more massive system ($8.9\pm2$ and $4.5\pm1.2 M_\odot$) at $1.5\pm0.5$~kpc 
and, also kinematically, a strong candidate-member of Cyg\,OB1. The 
more massive component is slightly evolved and appears to undergo 
non-radial $\beta~Cep$--type pulsations. The Doppler signal of the 
secondary is barely detectable. A more extensive study is important 
to fix masses more precisely, and an asteroseismological study would 
then become appropriate. Nevertheless, the position of the primary 
in the HR-diagram confines the age already reasonably well to 
$20\pm5$~Myr, indicating for Cyg\,OB1 a similar extent of star 
formation history as established for  Cyg\,OB2. 
\end{abstract}


\section{Introduction}

The star forming regions in the constellation Cygnus are among the most
active regions of star birth in our Galaxy. An overview of these regions
is presented in Reipurth \& Schneider (2008). These authors emphasized
the confusion of regions as near as hundreds of parsecs with regions at
1--2 kpc and even well beyond, due to the fact that in the direction of
Cygnus one looks down a spiral arm. Kinematical distances suffer from
the fact that the radial velocity gradient up to 4 kpc is smaller than
the typical velocity dispersion of interstellar gas. As a consequence,
even the need for the nine classical subgroups OB1 to OB9 in Cygnus is
still discussed, with for example a suggestion that OB1, OB8 and OB9
might just form one single subgroup (Melnik \& Efremov 1995). Hence,
distances derived from young eclipsing binary systems are of value to
localise the star forming regions and stellar associations.

A major study of binary stars in Cygnus has focused on Cyg~OB2, the
most famous and probably youngest subgroup. The Cyg~OB2 radial velocity
survey encompasses now twenty-five massive binaries (Kobulnicky et al.\
2012; Kiminki \& Kobulnicky 2012). Another very recent study (Mahy et
al.\ 2013) focus on bright O--type stars. 274 spectra taken over three
years reveal four spectroscopic binaries (three in Cyg~OB1 and one in
Cyg~OB9; none in Cyg~OB3 and Cyg~OB8). Earlier studies discuss binary
stars in open clusters superposed on Cyg~OB1. Zakirov (1999) presents
three eclipsing binaries with early--B primary components which are
plausible members of the open cluster IC~4996, but $UBVR$ photometry
without orbital-phase resolved spectroscopy did not lead to precise
fundamental parameters. Boeche et al.\ (2004) detect nine binaries out
of sixteen stars in the direction of NGC~6913, including seven binary
stars (one eclipsing) among the 12 presumed cluster members. They adopt
a distance of 1.6~kpc to NGC~6913, but emphasize (their Fig. 2) that
the cluster distance cannot be well constrained.
Very recently, 60 eclipsing systems in the field of the Cygnus OB7 
star forming region were identified by Wolk et al. (2013) from $JHK$ 
photometry in 100 nights spanning 1.5 year. A minority of them (23)  
are detached systems with orbital periods between $0\fd4$ and just  
over 13 days, and a majority (37) show continuous flux variations with   
periods ranging from 0.2 to 3 days. Most of the latter are thought  
to be W~UMa contact binaries. These stars have $J$ band magnitudes  
from 12.5 to 17.3, and are identified as diskless field stars. W~UMa  
type contact binaries are not observed in clusters younger than about  
1~Gyr. We did not find any literature on later-type binary stars   
unequivocally identified as members of the young associations.

The stars discussed in this paper are situated in the OB1--OB3--OB9
region. Two of them are suspected to be pre-main sequence systems with
A-type components, based on photometric studies (Zakirov \& Eshankulova
2005; 2006). They suggest the stars are connected to Cyg OB9 (V443 Cyg)
and to a group of OB-stars in the direction of Cyg OB1 (V456 Cyg).
Nelson (2011) points out the lack of spectroscopic knowledge for V456
Cyg and leaves open the question whether the primary component of this
system is an early--A or a late--A star. The third object, V2107 Cyg
(HD~191473), is an early-B system at the end of or slightly off the
main sequence with a primary component of luminosity class III to IV
(B0.5\,III, Roman 1951; B0\,IV, Morgan et al. 1955; B1\,III, Walborn
1971), and thus a particularly interesting age calibrator. Their
position in the Cygnus region, relative to the stellar subgroups, is
shown in Fig.~1.

\begin{figure}
\begin{center}
      \plotone{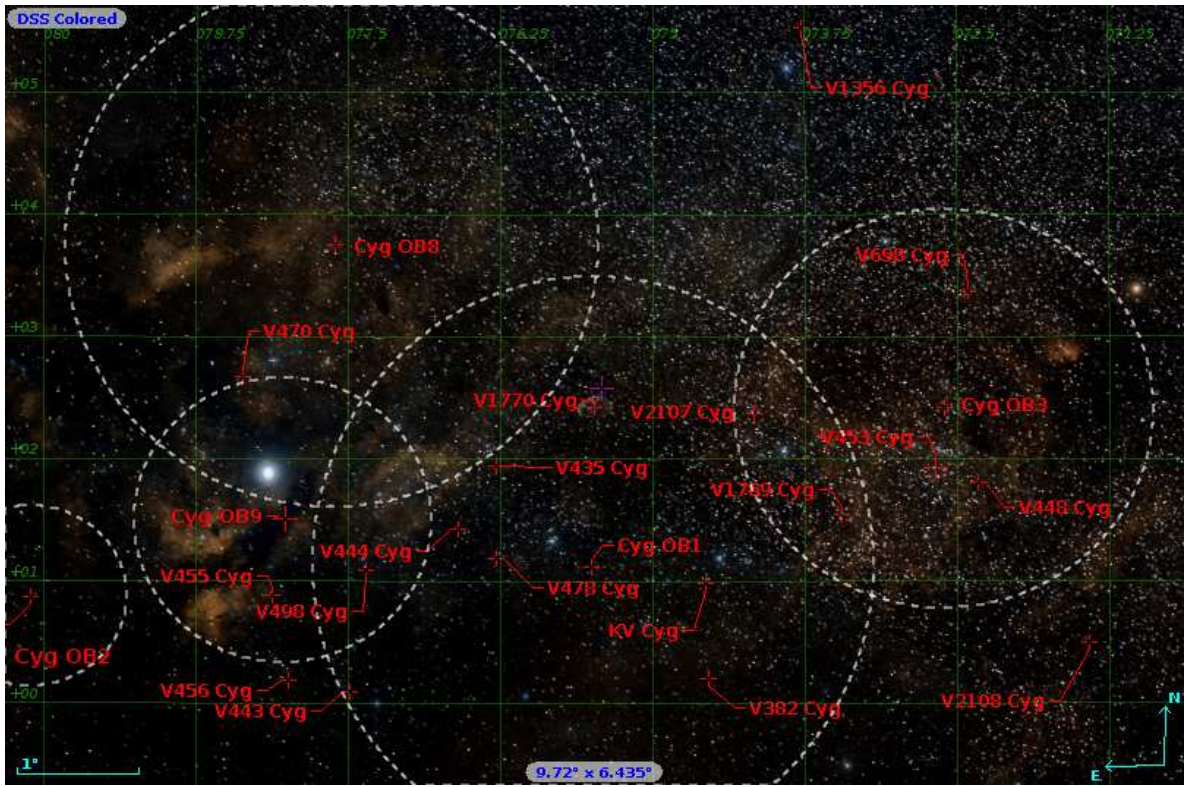}
\caption{Region of OB1,2,3,8,9 stellar associations in Cygnus according to
partitions of Blaha \& Humphreys (1989). \label{fig1}}
\end{center}
\end{figure}

In this paper, we summarize first the characteristics of the observations, 
and the analysis methodology applied (\S~2 \& 3). Thereafter, each of 
the three binary systems is discussed in turn (\S~4 to 6) and the 
general conclusions are presented in \S~7.

\section{Observations}

For each object, spectra at 11 to 15 orbital phases were obtained, and 
the light curve was covered completely in thousands of CCD frames obtained 
from 2009 to 2011 using various broad-band filters. The journal of the  
observations is given in Table~1 (spectroscopy) and Table~2 (photometry).   

\begin{deluxetable}{lcccrcccccccccccrcc}
\setlength{\tabcolsep}{0.05in} 
\scriptsize{}
\tablecaption{Journal of spectroscopic observations for program stars. Signal-to-noise (S/N) ratio refers to the continuum near 6500\,\AA.
\label{table1}}
\tablewidth{0pt}
\tablehead{
\colhead{No}& \colhead{HJD}      & \colhead{t$_{exp}$} & \colhead{S/N} & \colhead{$\phi$}  &\colhead{Z}& \colhead{HJD}      &
\colhead{t$_{exp}$}&\colhead{S/N} &\colhead{$\phi$}&\colhead{Z}&\colhead{HJD}&\colhead{t$_{exp}$}&\colhead{S/N}&
\colhead{$\phi$}&\colhead{Z} \\
            &\colhead{(-2400000)}&\colhead{(s)} & \colhead{}     & \colhead{($\phi$)} &\colhead{}       &\colhead{(-2400000)}&
\colhead{(s)}     &              &\colhead{($\phi$)}  & \colhead{}    &\colhead{(-2400000)}&\colhead{(s)}   & \colhead{}         &
\colhead{($\phi$)}  & \colhead{} 
}
\startdata
&\multicolumn{5}{c}{\textbf{V443~CYG}}&\multicolumn{5}{c}{\textbf{V456~CYG}}&\multicolumn{5}{c}{\textbf{V2107~CYG}}\\
1 &55455.3990&1200& 50 &0.477& 1.19  &55455.3562&1200& 155 & 0.291& 1.07  &55700.6345&3000& 150 & 0.008 & 1.18 \\
2 &55455.4181&1200& 50 &0.489& 1.27  &55455.3758&1200& 130 & 0.313& 1.11  &55790.6158&3600& 150 & 0.009 & 1.30 \\
3 &55455.4571&1200& 50 &0.512& 1.51  &55455.4330&1200& 120 & 0.377& 1.34  &55743.5015&3000& 150 & 0.013 & 1.21 \\
4 &55455.4751&1200& 50 &0.523& 1.67  &55455.4911&1200& 105 & 0.443& 1.84  &55161.2543&3600&  40 & 0.120 & 1.56 \\
5 &55456.3061&1200& 45 &0.023& 1.01  &55456.2268&1200& 130 & 0.268& 1.04  &55161.2978&3600&  40 & 0.130 & 2.11 \\
6 &55456.4785&1200& 30 &0.127& 1.74  &55456.3346&1200& 100 & 0.389& 1.04  &55162.1473&3600&  65 & 0.328 & 1.07 \\
7 &55456.4960&1200& 50 &0.137& 1.96  &55456.5170&1200& 110 & 0.594& 2.30  &55162.1908&3600&  50 & 0.338 & 1.20 \\
8 &55457.3371&1200& 40 &0.643& 1.05  &55456.5320&1200& 110 & 0.611& 2.65  &55775.6135&3000& 150 & 0.507 & 1.12 \\
9 &55457.3584&1200& 50 &0.656& 1.09  &55457.3141&1200& 80  & 0.488& 1.02  &55163.1388&3600&  60 & 0.560 & 1.06 \\
10&55457.4734&1200& 50 &0.725& 1.71  &55457.4390&1200& 90  & 0.628& 1.41  &55163.1822&3600&  60 & 0.570 & 1.18 \\
11&55457.5399&1200& 50 &0.765& 3.04  &55457.5106&1200& 95  & 0.709& 2.23  &55163.2256&3600&  60 & 0.580 & 1.39 \\
12&          &    &    &     &        &55795.4336&1800& 150 & 0.888& 1.09  &55164.1764&3600&  50 & 0.802 & 1.17 \\
13&          &    &    &     &        &          &    &     &      &        &55164.2199&3600&  55 & 0.812 & 1.37 \\
14&          &    &    &     &        &          &    &     &      &        &55164.2633&3600&  55 & 0.822 & 1.74 \\
15&          &    &    &     &        &          &    &     &      &        &55160.2510&3600&  50 & 0.886 & 1.51 \\
\enddata
%\tablenum{1}
\end{deluxetable}

\begin{deluxetable}{lcccc}
\setlength{\tabcolsep}{0.01in} 
\scriptsize{}
\tablecaption{Journal of photometric observations for program stars. \label{table2}}
\tablewidth{0pt}
\tablehead{
\colhead{Star}      & \colhead{Comparison/Check Star}       &  \colhead{Telescope(s) / Detector(s)} &
\colhead{Filter(s)} & \colhead{Number of Data} \\
\colhead{Mag., Colors} & \colhead{Mag.}                     &                             &    &
}
\startdata
\textbf{V443~Cyg}  &  GSC~3152~1079/GSC~3152~567  & 122-cm / Alta U42 & B & 409 \\
$V=$ 12.31  &         13.47 / 12.36        &                   & V & 460 \\
$B-V=$ 0.51 &                              &                   & R & 531 \\
$U-B=$ 0.05 &                              &                   &   &     \\
\textbf{V456~Cyg}  &  GSC~3152~491/GSC~3152~439   & 40-cm / Alta U47  & U & 1174 \\
$V=$ 10.80  &    11.12 / 10.53             &                   & B & 1200 \\
$B-V=$ 0.31 &    0.073 / 0.3               &                   & V & 1207 \\
$U-B=$ 0.11 &                              &                   & R & 1213 \\
            &        	                   &                   & I & 1204 \\
\textbf{V2107~Cyg} &  GSC~2683~1295/GSC~2683~191  & 40-cm - 30-cm / Alta U47, U42 - U47
& U & 5752 \\
$V=$ 8.63     &  11.19 / 9.51              &                   & B & 5885 \\
$B-V=$ 0.10   &   0.42 / 0.41              &                   & V & 6307 \\
$U-B=$ --0.60 &                            &                   & R & 7242 \\
\enddata
%\tablenum{2}
\end{deluxetable}

\subsection{Spectroscopy}

The spectra of V443~Cyg and V456~Cyg were taken with the Faint Object
Spectrograph and Camera ({\sc tfosc}) installed at the Cassegrain 
focus of the 1.5-m RTT telescope ({\sc rtt150}) of the T\"{U}B\.{I}TAK 
National Observatory (TUG), Turkey ($\sim 36^{\circ} 49^{m}$ N, 
$30^{\circ} 20^{m}$ E, altitude 2500-m). The highest resolution 
mode was used ($R(\lambda/\delta\lambda)\sim5100$). The list of grisms 
and the resulting resolutions can be found in the instrument's 
manual\footnote{\small\tt http://www.tug.tug.tubitak.gov.tr/rtt150$_-$tfosc.php}.  
The CCD chip has $2048\times2048$ pixels of $15\times15$ micron each. 
The wavelength range from 3300 to 9000 Angstr\"{o}m is covered completely in 11 
\'{e}chelle orders. All spectra were obtained in three observing 
sessions (September\,2010,  August\,2011, October\,2011).     

The spectroscopic observations of V2107~Cyg were performed at two sites.
The Coud\'{e}/\'{E}chelle Spectrograph ({\sc ces}) at the {\sc rtt150} 
telescope of TUG gives spectra with a resolving power of $R\sim40000$ 
and covers the wavelength range from 3960 to 8770 Angstr\"{o}m in 68 spectral orders 
on a SAO RAS A $1024\times1024$ nitrogen-cooled CCD camera (see Bikmaev 
et al. 2005 for technical details). The \'{e}chelle spectrograph {\sc hermes} 
(Raskin et al., 2011) at the 1.2--m Mercator telescope\footnote{\small\tt The Mercator 
Telescope is operated on the island of La Palma by the Flemish 
Community, at the Spanish Observatorio del Roque de los Muchachos of the 
Instituto de Astrofísica de Canarias.} was used in high-resolution mode 
($R \approx 80\,000$). It covers the wavelength 
range from 3700 to 9000 Angstr\"{o}m in 55 \'{e}chelle orders, with small gaps between 
the extreme-reddest orders. Five to six consecutive exposures of 600 seconds   
keep the Doppler shifts during a single exposure below 1 (2) kms$^{-1}$ for  
the primary (secondary), respectively, and summed provide a signal-to-noise  
ratio near 150 over a broad range. The data reduction pipeline of the  
spectrograph was used with the calibration frames available in standard   
operation mode, complemented by personally written programmes in the   
{\sc esomidas} package for continuum definition and removal of weaker  
cosmic rays by intercomparison of the subsequent exposures.

\subsection{Photometry} 

Photometry was performed at the Ulup{\i}nar Observatory of \c{C}anakkale 
Onsekiz Mart University ($\sim 40^{\circ} 06^{m}$ N, 
$26^{\circ} 30^{m}$ E, altitude 410-m).
The Schmidt-Cassegrain type telescopes and the Alta 
back-illuminated CCD detectors (Apogee $^{TM}$) were used. Bias and dark     
frame subtraction and flat field division was done using the {\sc iraf} 
platform. The {\sc C-Munipack}\footnote{\small\tt http://c-munipack.sourceforge.net/} 
software was used to perform aperture  
photometry with the size of the apertures three times the nightly FWHM  
values of the stellar images. Initially, several comparison and check  
stars were measured and those giving the lowest standard deviations in 
their differential magnitudes were selected for final differential 
photometry.

\section{Methodology}

Orbital periods are taken from the literature (V443~Cyg and V456~Cyg) 
or determined by combining our data with published measurements (V2107~Cyg) 
in order to gain precision by exploiting much longer time baselines than
covered by our own data alone. As a preliminary step, radial velocities   
were measured on the spectra for few individual lines, as far as line  
blending and component intensity allow. Together with published radial 
velocities, they were used with the purpose to obtain initial orbital  
parameters for the Fourier spectra disentangling code $\sc korel$  
(Hadrava, 1995) and $\sc fdbinary$ (Iliji\'{c}, 2003; see also
\footnote{\small\tt http://sail.zpf.fer.hr/fdbinary/fd3}). These codes solve 
for the pure Keplerian orbit, and deliver component spectra that are used to 
determine spectral types and, in particular, the temperature of the primary 
component. Thereafter, the light curves of the eclipsing systems are modeled, 
using the PHOEBE LC package (Pr\v{s}a \& Zwitter, 2005) which uses the  
Wilson-Devinney code (Wilson \& Devinney, 1971), inclusive the stellar  
atmosphere model (Kurucz, 1993) and a detailed reflection treatment (Wilson,  
1990). Circular orbits represent well the data for V443~Cyg and V456~Cyg. 
The orbit of V2107~Cyg is very slightly eccentric. The photometry puts 
a lower limit to $e \cos\omega$ and the spectroscopic solution must be  
tuned for consistency in the time of periastron passage (\S 6).
The mass ratio for the three systems was fixed to the value obtained 
from the spectroscopy. 

The light curve models used albedos $A_{1,2}$ and gravity darkening  
coefficients $g_{1,2}$ set to the generally adopted values (Rucinski, 1973;   
Lucy, 1967). Two-dimensional logarithmic limb-darkening coefficients were   
interpolated from the values given by van Hamme (1993). The temperature of   
the primary components were set to the spectroscopically estimated values.  
Semi-detached and detached configurations were tested. All three systems  
turn out to be well inside their Roche lobes.

\section{V443 Cyg}

The orbital period of V443~Cyg is $1\fd66220545$ (Zakirov \& Eshankulova, 
2005). The spectra disentangling codes converge to solutions with mass 
ratio close to unity (Table 3), both in the $H_\alpha$ and the $\lambda$4400 
regions. We used the jackknife method (see e.g. Bissell \& Ferguson 1975) 
to estimate the uncertainties and bias in the parameters, leaving out one 
spectrum at a time. It reveals that the size and orbital phase distribution 
of the present set of spectra is near the lower limit required for a 
robust orbital solution. For example, in the $\lambda$4400 region including 
H$\gamma$ and Mg\,{\sc ii}$\lambda$4481, we have nine instead of eleven 
useful spectra (the two additional ones being too noisy in the blue due 
to high airmass) and the orbit solution is instable in the sense that 
leaving out one more spectrum may change the mass ratio by 15 per cent. 
Thus, the precision of the orbit solution relies on the redder part of the 
spectrum. A significant gain in precision could be obtained by doubling 
the dataset. For example, allowing for (unexplained) subpixel shifts in  
two spectra that are least consistent with the whole set reduces the 
uncertainties by a factor of two. This gives an impression of the 
sensitivity of the present solution to data reduction or/and calibration 
issues (or to the assumption of the Doppler shifts reflecting a purely 
Keplerian orbit). The orbit derived from the spectra disentangling is given in Table~4.

The component spectra obtained in the two mentioned wavelength regions (Fig.~2)
indicate F-type components, in contrast to the expectation of Zakirov \& 
Eshankulova (2005) whose photometric analysis pointed to young A--type stars.  
The brightest component of V443\,Cyg is similar to HD\,128167, while the  
faintest component is similar to HD\,38393; but, apparently, $H_\alpha$ is  
partly filled-in with emission (Fig.~2). HD\,128167 ($\sigma$ Boo) is a F2V   
star. Most studies agreed on $T_{\mathrm eff} \approx 6700 {\mathrm K}$,  
$\log g \approx 4.3$ and [Fe/H] $\approx$ -0.4 dex (Edvardsson et al\. 1993;  
Gratton et al. 1996; Cunha et al. 2000; Cenarro et al. 2007). In view of its  
spectral type F2/3\,V, the effective temperature of the primary is fixed to   
6700~K in the light-curve analysis. HD\,38393 is classified F6.5\,V (Gray et  
al. 2006). Most studies agree on $T_{\mathrm eff} \approx 6350 {\mathrm K}$, 
$\log g \approx 4.3$ and [Fe/H] $\approx$ -0.1 dex (Edvardsson et al. 1993; 
Gray et al. 2006; Cenarro et al. 2007). Hence, V443\,Cyg contains an early-F 
and a mid/late-F star. 

\begin{figure}
\begin{center}
\plotfiddle{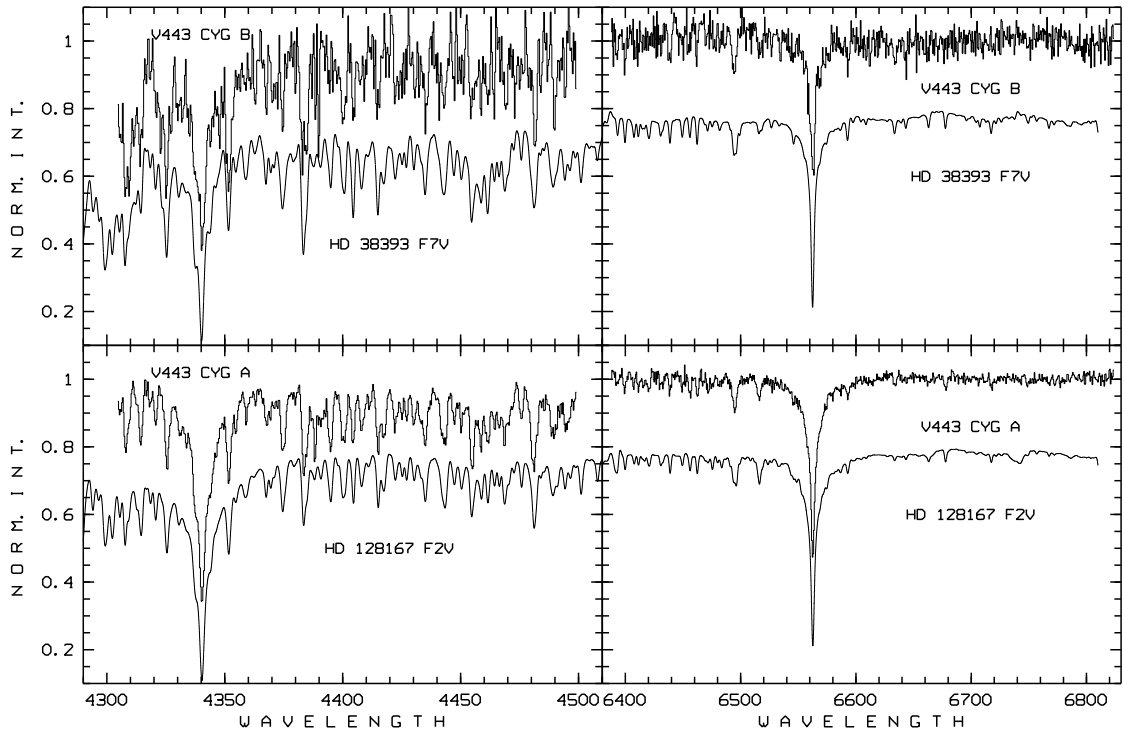}{2cm}{-90}{60.}{60.}{-250}{325} \\
\caption{Reconstructed component spectra of V443~Cyg, compared to spectra of
bright stars in the {\sc eso uves pop} library degraded to the resolution of the 
{\sc tfosc} spectra of V443~Cyg.}\label{fig2}
\end{center}
\end{figure}

Fig.~7 shows the best-fitting LC models and Table~5 the associated parameters. 
Table~6 lists the characteristics of the components and the orbit, as 
derived from the combined spectroscopic-photometric analysis. The masses and  
radii are, within the substantial uncertainties, compatible with the spectral   
types. The system is significantly less massive than the $3.2 + 2.5 M_\odot$   
mentioned by Zakirov \& Eshankulova (2005), and has a mass ratio much nearer   
to unity. Component masses are of the order of 1.1 to 1.2 $M_\odot$. The  
difference in spectral type and temperature is better compatible with a mass   
ratio around 0.9 (that is within the error bars of the present analysis) than  
with the best nominal value $M_2$/$M_1$~=~0.98 (see Fig.~8). For such a    
light system still near the zero-age main sequence, the exact age is largely  
uncertain. Isochrones derived from the tracks of Bertelli et al. (2009), where initial 
rotation velocity for tracks and isochrones is not taken into account, show  
that the system might be either as young as 2$\times10^7$ years or more than one 
Gy old. The system is at a distance of $0.6 \pm 0.2$~kpc.

\clearpage

\section{V456 Cyg}

In accordance with Nelson (2011), we used an orbital period of 
$0\fd89119559$. Zakirov \& Eshankulova (2006) suggested a parabolic trend   
in the O--C diagram of V456~Cyg. However, this assumption leads to less 
consistency between our light and RV data that were obtained at different 
epochs. Moreover, we notice that the more recent orbital period published 
in Hoffman et al. (2008), $P_{\mathrm NSVS} = 0\fd61634$, based on the 
NSVS photometry (Wozniak, Vestrand, Akerlof et al. 2004), is an alias of 
the true period, $P^{-1}_{\mathrm NSVS} - P^{-1}_{\mathrm true} = 0.5004$.  
It is in disagreement with the extensive list of eclipse timings in the 
literature. This is also the case for several other stars in that paper.

The RVs given by Nelson (2011), shifted by $-4.6$~kms$^{-1}$, were used 
together with ours to determine the initial parameters for the spectra 
disentangling codes. This shift was determined comparing for both data 
sets the linear relation between the RVs of both components, with the 
gradient giving the mass ratio $q$ (identical for both sets) and the 
zero-point defined by $(1+q) V_{\gamma}$. While not stated explicitly in 
his paper, we suspect that the velocities of Nelson, by the use of templates, 
are relative to the systemic velocity of the system rather than on an 
absolute velocity scale. The applied shift is estimated to be uncertain by 
1.5~km/s.

The orbit derived from the spectra disentangling is given in Table~4. When 
analysed separately, the Nelson RVs give 5 percent lower velocity amplitudes. 
Therefore, the formal uncertainties (1 percent) on $K_1$ and $K_2$ should be 
interpreted with caution. Uncertainties of 2.5 percent on the masses may be 
more realistic.

The reconstructed component spectra of V456~Cyg (Fig.~3) definitely rule 
out `Model 1' of Nelson (2011). The hydrogen lines of the primary are 
consistent with late--A type stars, with $H_\alpha$ suggesting A9 and
H$\gamma$ rather A7. The Balmer lines do not show the strongly Stark
broadened wings of early--A type stars, and the dynamical analysis does
not leave space to consider higher luminosity. However, the ratios of 
$\lambda$4417 (Fe\,{\sc ii} and Ti\,{\sc ii} blend) and Ti\,{\sc ii}\,$\lambda$4444 to Mg\,{\sc ii}\,$\lambda$4481 suggest A4, 
and Ca\,{\sc i}\,$\lambda$4227 and the Ca\,{\sc ii}\,K line (the latter
not disentangled, but when the secondary component is eclipsed) suggest
A2. Thus, one is tempted to classify V456~Cyg~A as a slightly metal-weak
star, A2hA7mA4. Its rotation velocity is $100 \pm 10$~kms$^{-1}$. With
the hydrogen lines being the most sensitive temperature parameter, the
temperature of V456~Cyg~A is fixed at $7750 \pm 150$\,K for the light
curve analysis.

\begin{figure}
\begin{center}
\plotfiddle{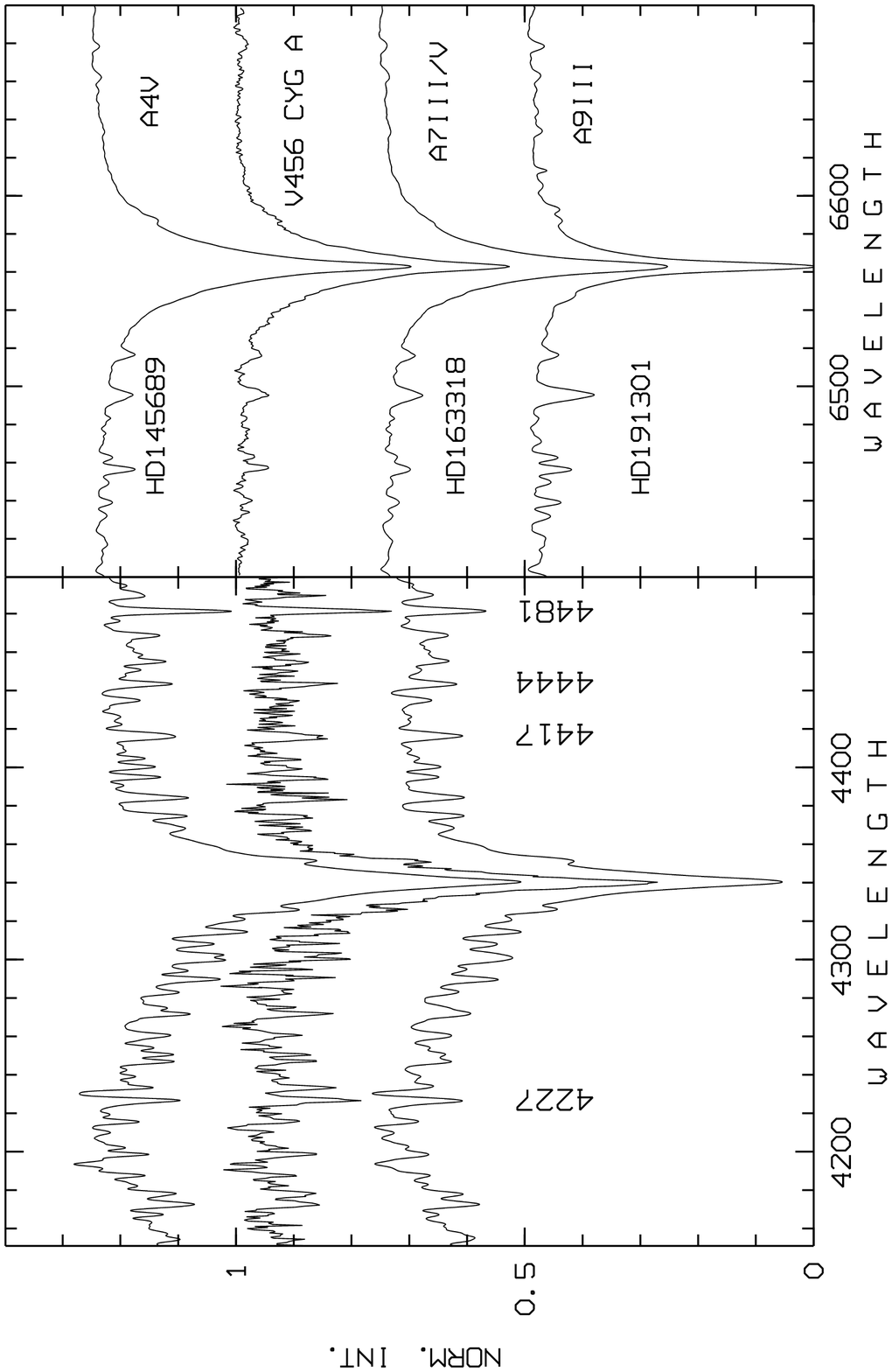}{5cm}{-90}{45.}{45.}{-200}{300} \\
\plotfiddle{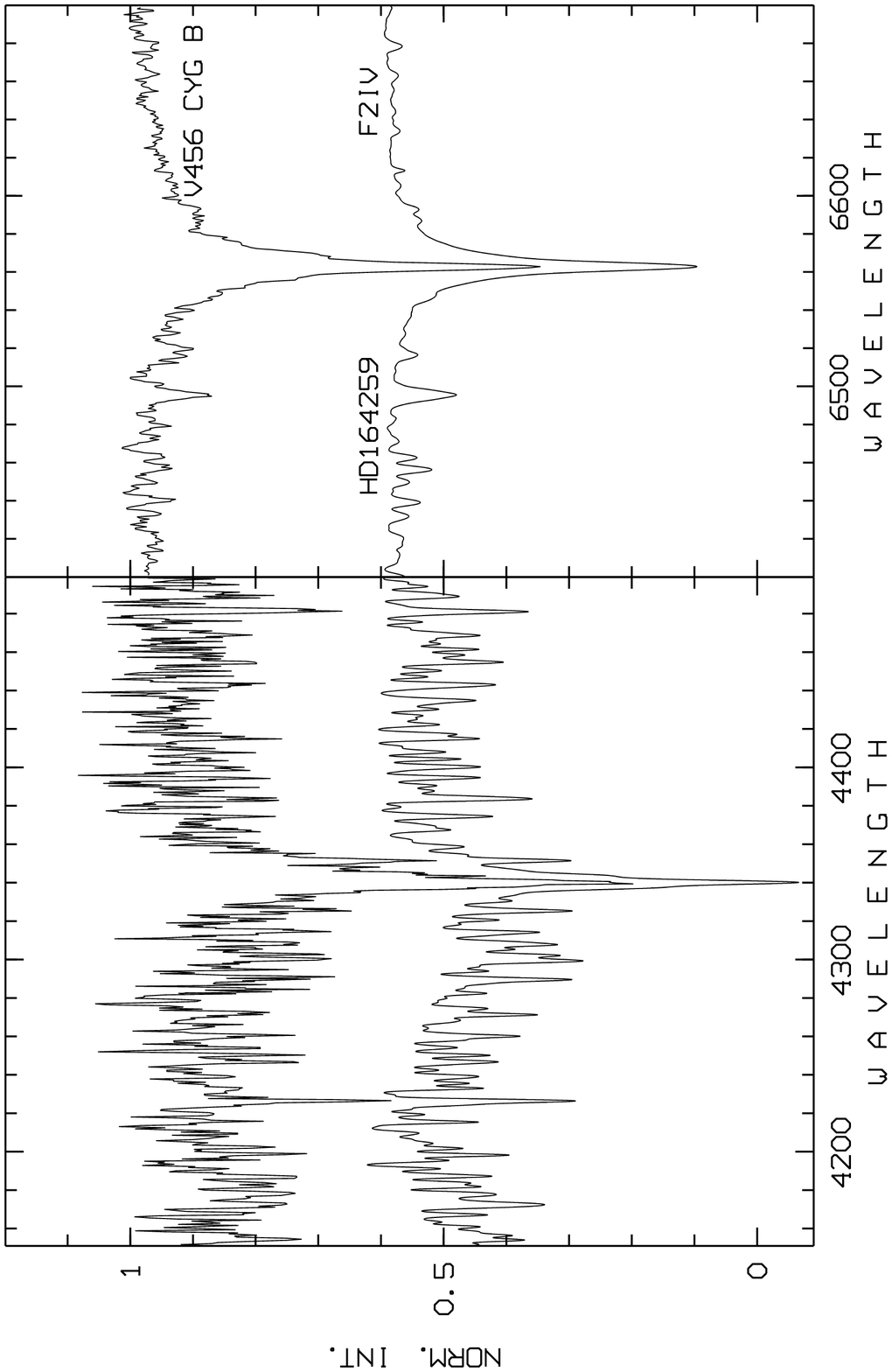}{5cm}{-90}{45.}{45.}{-200}{225} \\
\caption{\textit{Upper:} Parts of the disentangled spectrum of V456\,Cyg\,A compared to {\sc uves pop}
spectra of A--type stars degraded to the resolution of the {\sc tfosc} 
instrument. The {\sc uves pop} spectra were `normalised' tracing a linear 
approximation to the continuum in each of the regions. Velocity shifts were 
applied to all spectra to ease comparison of spectral lines. \textit{Lower:} Parts of the disentangled 
spectrum of V456\,Cyg\,B compared to an early-F star in the {\sc uves pop} library.}\label{fig3}
\end{center}
\end{figure}

The Mg\,{\sc ii}\,$\lambda$4481 line is stronger than Fe\,{\sc i}\,$\lambda$4383, and the Mg\,{\sc ii}
line does not show a greater weakening than the general metallic-line
spectrum with respect to the hydrogen lines. As both of these criteria
are contrary to what is seen in {\it most} $\lambda$ Bootis stars (see e.g.
Gray 1988), the star very probably does not belong to this class. Whether
the hydrogen lines are in a subtle way peculiar, as seen in $\lambda$ Bootis
type stars, is not easily evaluated on the reconstructed spectrum.

The secondary component rotates significantly slower, but a precise
estimate is hard at the {\sc tfosc} resolution. We put an upper limit
of 50 kms$^{-1}$ to $v \sin i$. The star shows a general resemblance with
the degraded high-resolution {\sc uves} spectra of HD\,164259, a F2{\sc iv}
star, although not perfectly. The noise level in the reconstructed spectrum
of the fainter companion and the restricted resolution do not permit to
confirm independently a slightly low metallicity as seen in the primary.

The system certainly deserves a more detailed analysis at higher resolution.
In conclusion, the reconstructed spectra show a system consisting of a
late--A and early--F spectrum, with at least the primary being somewhat
metal-weak.

Fig.~7 shows the best-fitting LC models and Table~5 the associated 
parameters. Table~6 lists the characteristics of the components and  
the orbit, as derived from the combined spectroscopic-photometric 
analysis. Our solution is near to Nelson's (2011) `Model 2', but in 
some aspects different. Nelson's problem of disparate ages related 
to the radius of the less massive star and its $\log g$ being larger 
than for the primary does not survive in our solution ($1.68 + 1.47 R_\odot$; 
$\log g = 4.26 + 4.30$). We find also masses 8\% higher (while agreeing 
on the mass ratio), and the luminosity of the primary exceeds Nelson's by 
30\%. The system is at a distance of $0.50 \pm 0.03$~kpc.

With the precision obtained for this system, a discussion of its position  
in the Hertzsprung-Russell diagram is interesting. The stars are slightly  
too cool and too small for their masses, unless a metallicity as high as 
$Z = 0.035$ is assumed. The least we can state is that the photosphere 
does not provide arguments for an enhanced metallicity. Therefore, it 
appears worthwhile to investigate the option that the system might be young, 
near to but still on its way to the main-sequence. A comparison with the 
tracks of Bertelli et al. (2009) shows an encouraging agreement (see Fig.~8), 
although the less massive component is predicted further from the main-sequence 
than observed. In order to discern whether the system is arriving at the zero age main sequence (25 Myr),  or older 
(315 Myr) with higher metallicity, more detailed spectroscopy is required. Any age between the quoted extremes is presently still an option.

\section{V2107~Cyg}
Phasing our photometry together with the {\sc hipparcos} satellite 
photometry, the orbital period was fixed to $4\fd2846$. An earlier 
spectroscopic study (Mercier 1957, further M57) suggested an orbit 
with small eccentricity ($e = 0.045$). Lucy \& Sweeney (1971), 
further LS71, in a paper rejecting low-eccentricity orbits for many 
systems after considering the bias in the determination of $e$ 
(towards larger $e$ for low $e$), confirmed that the orbit of 
V2107~Cyg is probably not circular ($p = 0.011$) and updated the 
orbital parameters, inclusive $e = 0.070$. Therefore, a discussion  
of the reality of a non-circular orbit despite the 4--day period   
is presented in \S~6.2.

Since the secondary star is occulted completely during six hours 
around phase 0.5, the spectrum of the hotter component (A) is 
observed directly. The spectrum of the secondary is not readily   
visible in any of the observed spectra, even after (scaled) 
subtraction of the observed pure component--A spectrum. The light  
curve suggests a contribution of the secondary to the total light  
of only 6 per cent. Hence, a large range of $K_2$ must be explored  
in the spectral disentangling. A further complication in identifying 
the signal of the secondary is posed by the line profile variability 
in the primary. 

Hence, we discuss in subsequent subsections first the spectrum of 
V2107\,Cyg~A and the character and cause of the variable metal line 
profiles. Then, we evaluate the need for a non-circular orbit. And 
finally, the orbit and the fundamental parameters of the stars are 
presented.

\subsection{The spectrum of V2107\,Cyg~A}

The spectrum of the hotter component (A) is observed directly near 
phase 0.5. Comparison with pure rotation profiles (Gray, 1992, 
$\epsilon = 0.6$) for ten sufficiently unblended metal lines between  
4260 and 5740 Angstr\"{o}m, mainly due to Si\,{\sc iii} and O\,{\sc ii}, suggests   
an apparent rotation velocity of $v \sin i = 84 \pm 4$ kms$^{-1}$. 
This value should be interpreted as an upper limit to $v \sin i$ as 
other velocity fields (macroturbulence, pulsations, etc.) were not 
considered. A comparison of hydrogen and helium lines with model 
calculations (LTE line-blanketed models with depth dependent non-LTE 
line formation computations, as described in Vrancken et al., 1997),
fixing $\log g = 3.6$ cgs as derived from the analysis of the binary, 
suggests $T_{eff} = 22500 \pm 1500$~K. The uncertainty takes into 
account that better models are required as lines or line cores 
formed in higher photospheric layers are inadequately represented  
(Fig.~4). Such detailed analysis is out of the scope of this paper.  

\begin{figure}
\begin{center}
\plotfiddle{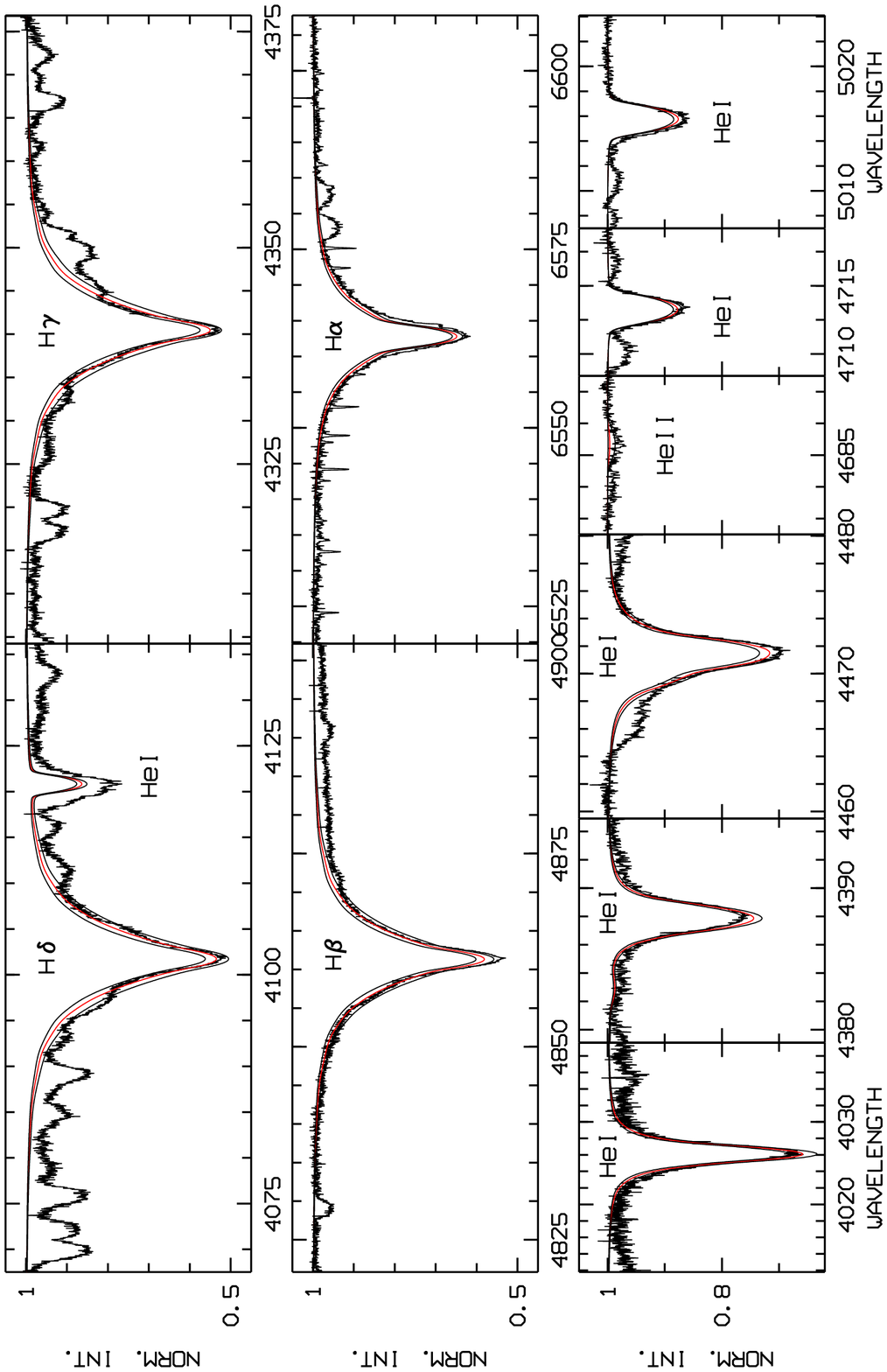}{5cm}{-90}{60.}{60.}{-280}{300}
\caption{Hydrogen and helium lines in total eclipse of V2107~Cyg
(pure spectrum of the primary) compared to models for
$\log g = 3.6$ cgs (see Table~6) and effective temperatures of 20\,000 (black line),
22\,000 (red line) and 24\,000~K (black line). The red line is the fit for the intermediate temperature. Deeper absorption in the models
corresponds to lower temperature, except for the He\,{\sc ii} line. Notice that He\,{\sc i}$\lambda$4121
is heavily blended by metal lines, while the red wing of H$_{\beta}$ is affected by a diffuse interstellar
band.}\label{fig4}
\end{center}
\end{figure}

Most metal lines show asymmetric line cores. The high-resolution    
spectra observed at three different epochs in primary eclipse 
(Fig.~5) confirm that V2107\,Cyg~A is an intrinsic spectrum variable.     
Since these spectra were taken at essentially the same orbital phase,   
the origin of the variability is not due to the binary character.   
With local changes in line depth of up to four percent, the origin   
must be in the primary component rather than in the secondary that  
contributes only six percent to the total flux.   

\begin{figure}
\begin{center}
\epsscale{0.7}
\plotone{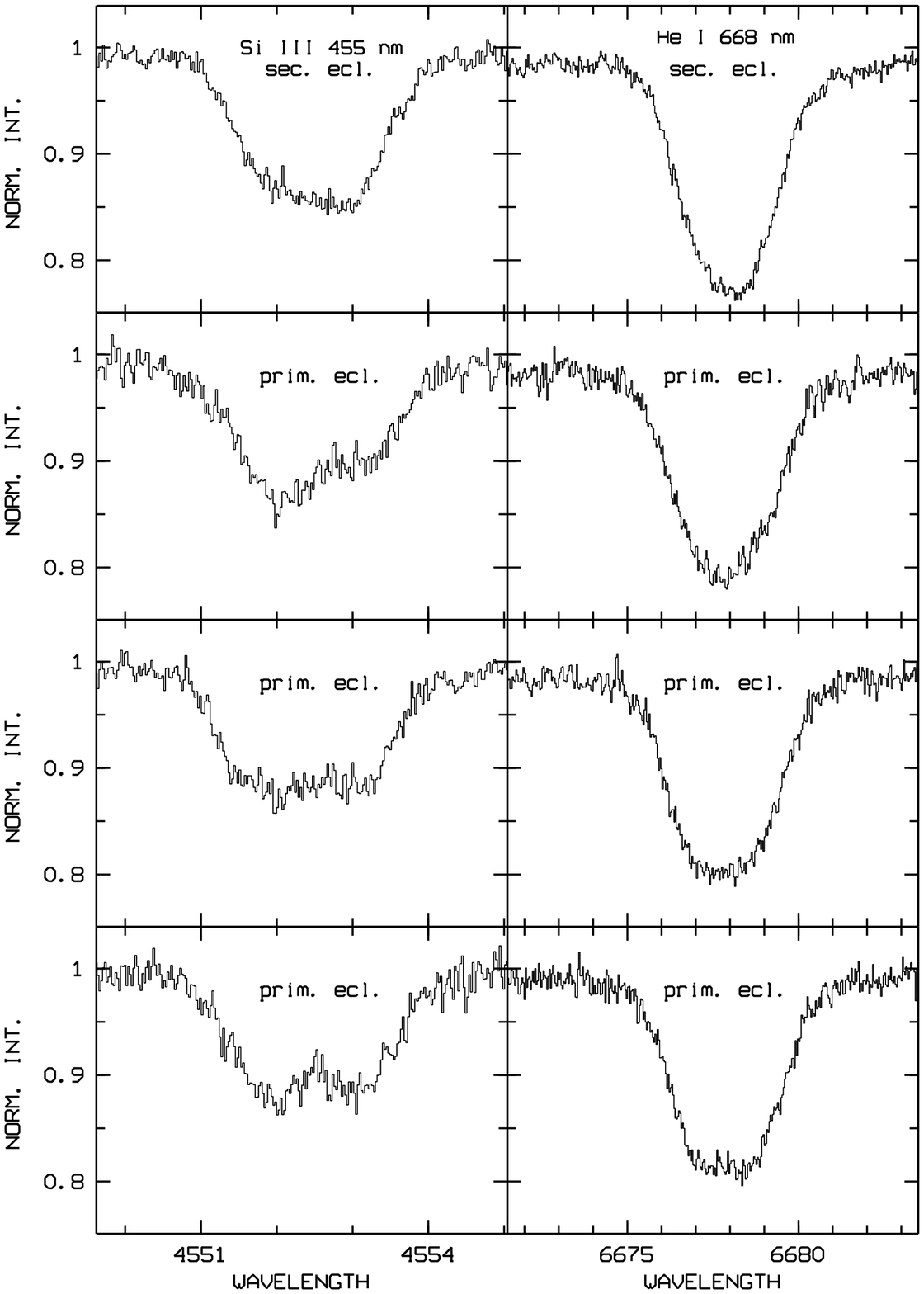}
\caption{Line variability seen in two lines. The upper panel shows the spectrum 
of V2107~Cyg\,A (mid-secondary eclipse). The three lower panels show 
primary eclipse spectra taken during three different orbital cycles. 
In each case 5 to 6 subsequent 10-minute exposures were added after 
small shifts to correct for the orbit.}\label{fig5}
\end{center}
\end{figure}

Temperature and gravity place V2107~Cyg~A in the $\beta$~Cephei
instability strip. The observed type of line profile variations
suggests strongly it is a non-radial pulsator. The photometry   
supports this viewpoint: the residuals of the orbital light curve  
show evidence for short-term light variability (Fig.~6). In order  
to enhance the sensitivity for low-amplitude variations, the  
check was performed on the highest-quality data, with estimated  
rms uncertainties of 0.005 to 0.008 mag. The selected data set  
contains 2269 measurements obtained in 15 nights in a time interval  
of two years, with nightly series covering 2.2 to 6.7 hours. Data  
in ingress and egress of eclipses were avoided.

\begin{figure}
\begin{center}
\plotfiddle{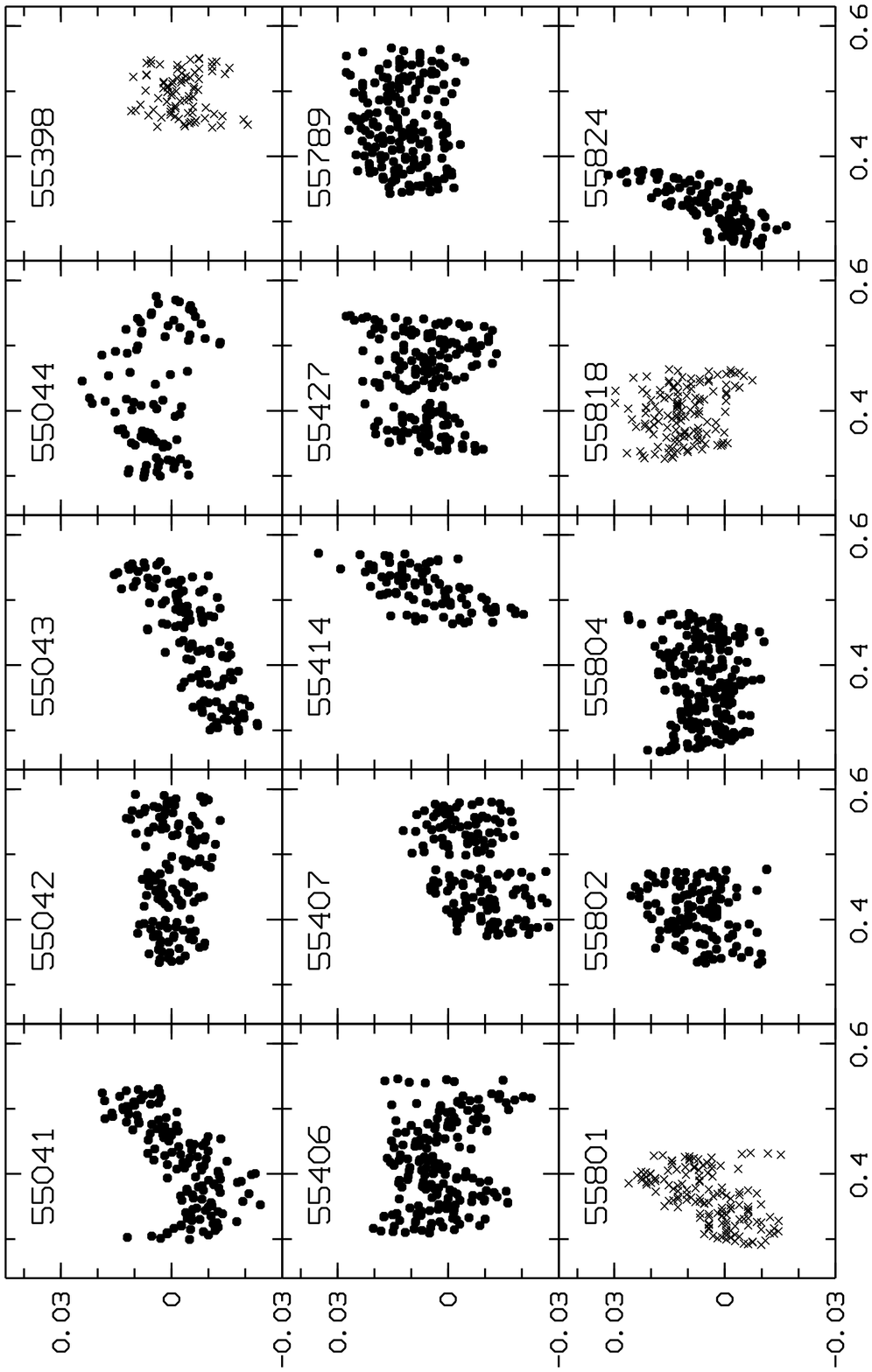}{5.2cm}{-90}{45.}{53.}{-225}{260}
\caption{Photometric variability of residuals from the LC solution of V2107~Cyg. 
Number in each panel is Julian Day -- 2400000.}\label{fig6}
\end{center}
\end{figure}

The photometric data are suggestive of short-term light variability,
either multi-periodic or irregular. A period search did not single 
out any dominant frequency, although several peaks (many of them 
aliases of each other) are above the noise level. Test computations 
with artificial data and combinations of two frequencies with peaks 
in the observed power diagram indicate that the noise level and the 
observed time windows may be insufficient to retrieve the true 
frequencies uniquely. Nevertheless, the observed light and spectrum 
variability should encourage larger-scale observational efforts. 

\subsection{Circular orbit?}

M57 determined the spectroscopic orbit from radial velocity  
measurements on 16 plates obtained between July 27 and November 23,   
1955. Depending on the plate, 4 to 6 line positions were measured    
(He\,{\sc i}\,$\lambda$4009.27, $\lambda$4120.81, $\lambda$4026.19, 
$\lambda$4387.93 and $H\gamma$ and $H\delta$). He notes that only He\,{\sc i}\,$\lambda$4009.27  
and $\lambda$4387.93 are of good quality to measure RVs. Table~3 
compares his orbit solution to others derived from the same data 
and discussed here. All these solutions used $P4\fd2876$ as in  
M57, but it was checked that updating the orbital period to the 
value we derived ($4\fd2846$) has negligible influence.  

\begin{deluxetable}{llllll}
\setlength{\tabcolsep}{0.05in} 
\scriptsize{}
\tablecaption{Orbital solutions of V2107\,Cyg based on the M57 data. LS71 
refers to Lucy \& Sweeney (1971), the last three column contain our 
computations (least-squares eccentric orbit, least-squares circular orbit, 
jackknife bias-corrected orbit). \label{table3}}
%\tablewidth{0pt}
\tablehead{
\colhead{Parameter} & \colhead{M57} & \colhead{LS71} & \colhead{ecc} & \colhead{cir} & \colhead{jack} \\ 
}
\startdata
$K$ (km/s) & 106.5 & 106.8  & 107.3$\pm$1.6 & 104.6 & 106.0$\pm$4.0 \\ 
$e$        & 0.045 & 0.070 & 0.073$\pm$0.015 & 0 & 0.06$\pm$0.03 \\
$\omega (\deg)$ & 134 & 122 & 121$\pm$14 &  & 115$\pm$17 \\
$T_{\mathbf{p}} - 2\,435\,400$ & 30.78 & 29.158 & 30.60$\pm$0.01 & & 30.53$\pm$0.20 \\
$V_{\mathbf{sys}}$ (km/s) & -3.7 & -3.8 & -3.8 & -3.9 &  \\
$f(M)$ ($M_{\odot}$) & 0.53 & 0.54 & 0.54 & 0.51 & 0.53 \\
$\sigma(\mathbf{O-C})$ (km/s) & 4.8 & 5.0 & 4.3 & 6.5 &  \\
\enddata
\end{deluxetable}

Computing the least-squares eccentric and circular orbits, we 
confirm the low probability that the orbit is circular ($p = 0.011$)  
and the LS71 orbital parameters, except their deviating value for 
the time of periastron passage. The uncertainties mentioned in our 
solution are computed with the formalism developed in LS71. In view 
of the evidence for non-radial pulsations that would influence the 
line position measurements in a non-random way, we used also the 
jackknife method to estimate bias in the orbital parameters and  
uncertainties. The bias does not exceed the 1-$\sigma$ level, but, 
more importantly, the jackknife method suggests larger uncertainties, 
by a factor of about 2, for $K$ and $e$. Adopting these uncertainties 
suggest not to reject a circular orbit (if we use the LS71 criterion 
$p > 0.05$). Hence, the M57 data may not be sufficient to reject a 
circular orbit. Using our set of spectroscopic data does not lead to 
a stronger conclusion.

However, the light curve provides a completely new piece of evidence.
Fitting a circular orbit leaves in ingress and egress of the secondary
eclipse systematic residuals that can be removed for a phase separation
between the eclipses that is slightly different from half a period, in 
the sense that the secondary eclipses seen in the $BVR$ data occur 
slightly earlier than half a period after primary eclipse, requiring  
$e \cos\omega \ge 0.015$. Notice that the Hipparcos data indicate an 
opposite effect, if any. However, these data are less straightforward 
to interprete, both because the eclipses are sparsely covered (from 
observations in different orbital cycles), and because a significant 
part of the data in eclipse tends to have larger than average uncertainty 
estimates in the Hipparcos catalogue. Taken at face value, a varying  
time separation between the eclipses might indicate motion of the line 
of apsides.

Therefore, we searched for solutions that would represent as well our 
photometric data, as the Hipparcos ones, and the Mercier spectroscopy. 
The search was made for $0.015 \le e \le 0.07$ taking into account that 
$\omega$ may change with time. Towards the larger eccentricities in 
this interval, the $\omega$ value at the time of the recent photometry 
is nearer to (as compared with the M57 epoch) but larger than 90 degrees.

However, we were unable to constrain the eccentricity in this way, as 
the Hipparcos photometry requires $\omega$ somewhat below 90 degrees, 
but not that much that almost one full apsidal motion period would have 
elapsed from M57 to the present time. Higher-precision data are needed 
to pinpoint the small eccentricity. From the available data we conclude 
$e = 0.045 \pm 0.03$. In order to show the need for a non-circular orbit,
Fig. 7, lower panel, shows the solution for $e = 0.045$ and 
$\omega = 102 \deg$, as well as for a circular orbit.

Using the $H_\alpha$ region of our spectroscopy, we do not find support 
for combinations ($e$, $\omega$) as derived from photometry for 
$e \ge 0.03$, in the sense that such combinations do not lead to 
solutions with lower chi-square than with a circular orbit. However, 
in view of the restricted number of spectra and their phase distribution 
we prefer not to use this argument to exclude firmly slightly higher 
eccentricity. Fortunately, $K_1$ depends on the assumed value $e$ only 
at the $\pm 1$~km\,s$^{-1}$ level. The value of $K_1$ is consistent with 
radial velocity measurements obtained directly from specific lines 
(but this classical method leads to larger uncertainty in $K_1$).

The dependency of $K_2$ on $e$ is negligible in view of the much larger 
uncertainty associated with the fact that the signature of the Doppler 
movement of the faint secondary component is detected only marginally.
The signal-to-noise ratio of the $\sc{ces}$ spectroscopy is  
insufficient to reconstruct any specific feature in the spectrum of 
the secondary. Only the integrated Doppler signal is detected. In 
view of the marginal detection, an effort was made to map the 
multi-dimensional chi-square space in the neighbourhood of the 
deepest minima, for a circular orbit (Fig.~8) . Test computations  
for slightly eccentric orbits shows that details of the geometry 
in chi-square space change with $e$, but not the global structure.

The detailed gridding reveals the existence of various chi-square
minima similar in depth for $185 \le K_2 \le 230$~km\,s$^{-1}$, while 
$K_1$ and the phase of maximum velocity are well-determined. We 
constructed the grid up to far outside of this $K_2$ range, for 
$K_2 > K_1$, but found no similar minima outside the range quoted 
above. The most prominent minima are suspiciously narrow, in view 
of the width of the core of the $H_\alpha$ line, and possibly linked to 
structure in the noise or in the spectra (telluric lines, pulsation?). 
Therefore, the right-hand side panel of Fig.~8 presents a smoothed 
version of the chi-square space. The uncertainty quoted for the final 
spectroscopic orbit (Table 4) and the fundamental parameters of the 
components (Table 6) take into account the location of the multiple
minima, rather than selecting quite arbitrary one of the narrow minima.

\begin{deluxetable}{lcccc}
\setlength{\tabcolsep}{0.05in} 
\scriptsize{}
\tablecaption{Spectroscopic orbital parameters of program stars. The numbers between brackets behind the value of the 
period are the uncertainties in the last digits of $P$. \label{table4}}
%\tablewidth{0pt}
\tablehead{
\colhead{Parameter}     & \colhead{V443~Cyg}   & \colhead{V456~Cyg} & \multicolumn{2}{c}{V2107~Cyg}           \\
                        &                      &                    & \colhead{Sol.1}      & \colhead{Sol.2}
}
\startdata
$P$ (days)    & \multicolumn{1}{c}{1.66220545(43)} & \multicolumn{1}{c}{0.89119559(17)}  & \multicolumn{2}{c}{4.2845923(25)} \\
$T_{0}$ $(HJD-2450000)$ & 5456.2677(0.0025)& 5455.0966(0.0016)& \multicolumn{2}{c}{5259.3024(0.0004)} \\
$K_{1}$ (km s$^{-1}$) &   117.4 (6.2)    & 152.1  (1.7)     & \multicolumn{2}{c}{104 (1.5)} \\
$K_{2}$ (km s$^{-1}$) &   120.2 (4.2)    & 179.4  (1.7)     & 187 (5) & 207(20)\\
$e$ [fixed]           &     0.0          &   0.0            & \multicolumn{2}{c}{$<$0.03} \\
$V_\gamma$ (km s$^{-1}$)&  -19.7 (0.7)    &  -4.2  (1.0)     & \multicolumn{2}{c}{-6.3 (0.2)} \\
$q (M_2/M_1)$         &     0.977(0.067) &   0.848(0.018)   & 0.56(0.02) & 0.50(0.06)\\
\enddata
\end{deluxetable}

\begin{figure}
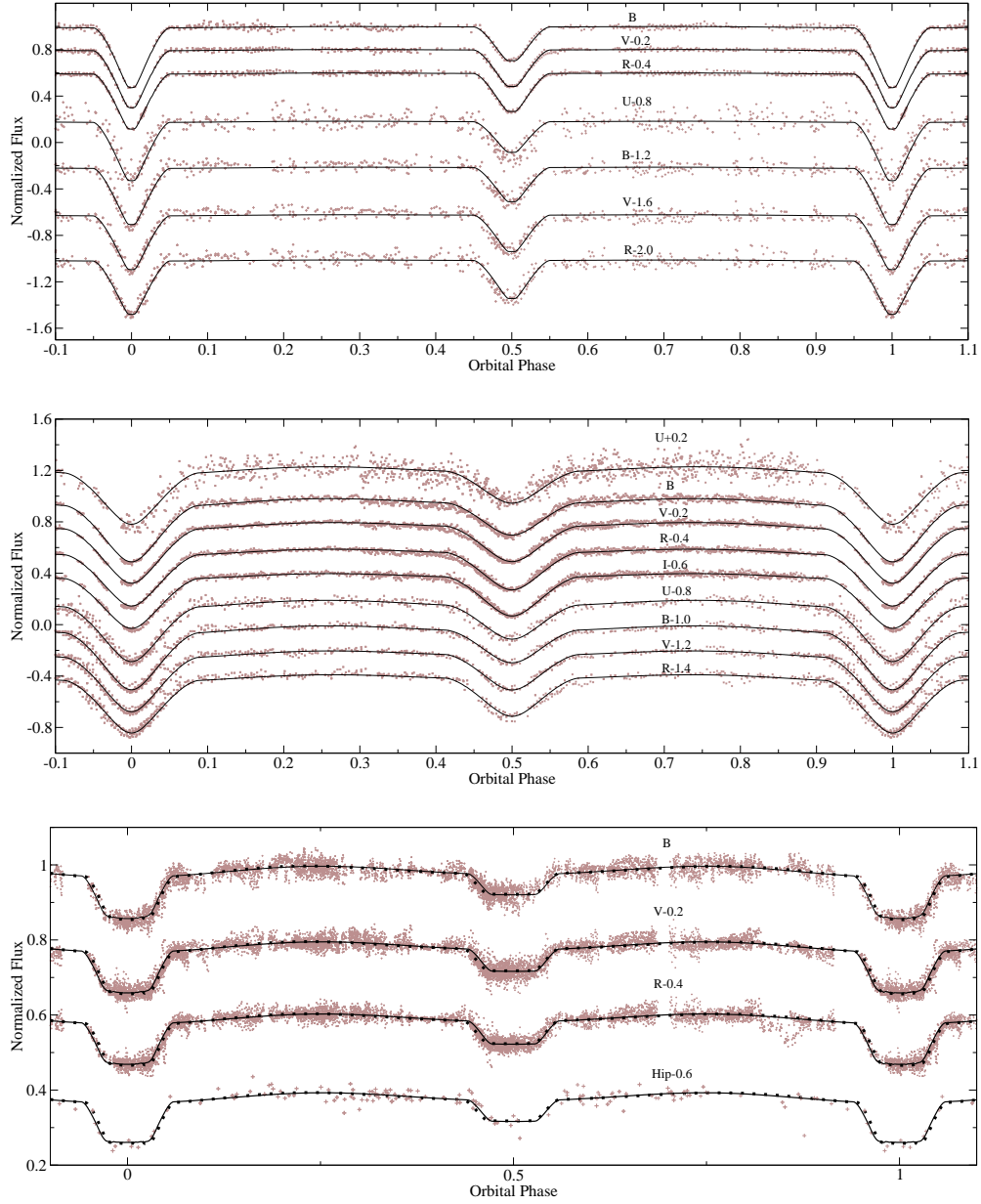

\begin{center}
\epsscale{0.8}
\plotone{fig7a.eps} \\
\vspace{5mm}
\plotone{fig7b.eps} \\
\vspace{5mm}
\plotone{fig7c.eps} \\
\caption{Best fitting LC models \textit{(solid lines)} for V443~Cyg
\textit{(top)}, V456~Cyg \textit{(middle)} and V2107~Cyg \textit{(bottom)}. For
V443~Cyg and V456~Cyg, LC data belonging to present study and
Zakirov \& Eshankulova (2005, 2006) are shown with cross and plus signs,
respectively. Eccentric orbit solution for V2107~Cyg is shown with black dots.}\label{fig7}
\end{center}
\end{figure}

\clearpage

\begin{figure}
\begin{center}
%\epsscale{0.7}
\plotfiddle{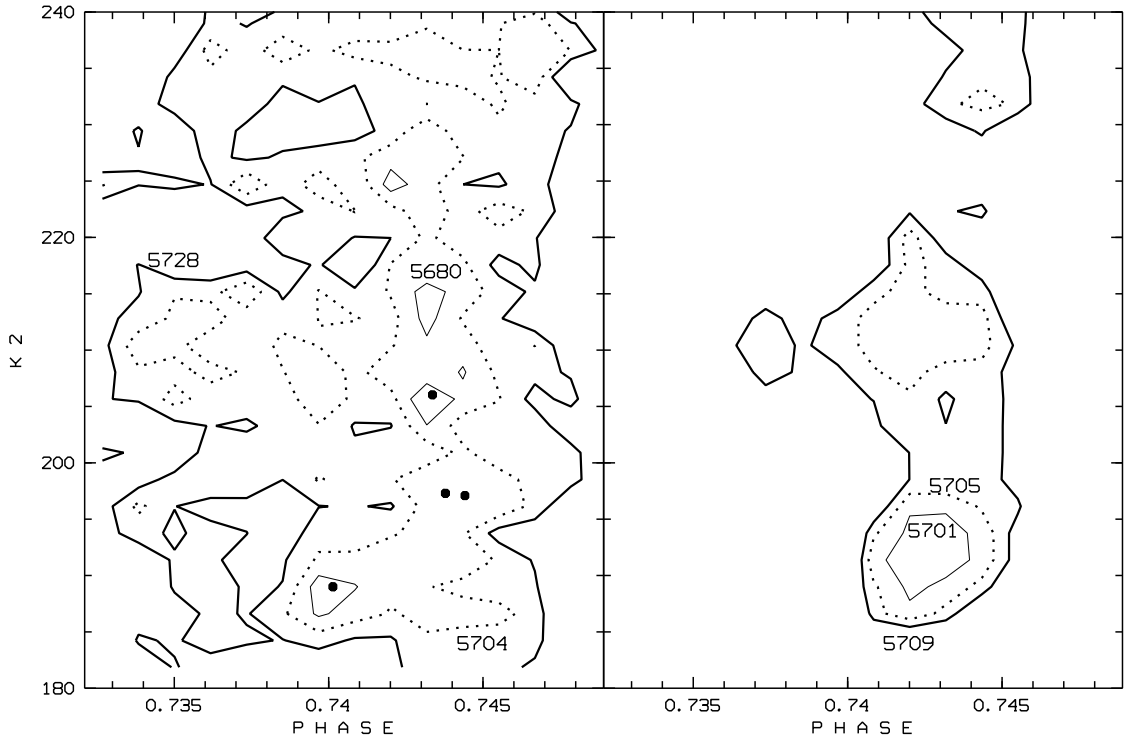}{5cm}{-90}{60.}{60.}{-280}{300}
\caption{$\chi^2$ map for the $H_\alpha$ region of V2107 Cyg, as 
seen in the plane ($\phi$, $K_2$), without (left panel) and 
with median filtering (right panel) to disregard very narrow 
$\chi^2$ structure. See explanation in \S~6.2.} \label{fig8}
\end{center}
\end{figure}

\clearpage

\begin{deluxetable}{lccc}
\setlength{\tabcolsep}{10pt} 
\scriptsize{}
\tablecaption{Results from the simultaneous solution of LCs of program stars in
several filters. Adjusted and fixed parameters are presented in separate panels of the
table. Uncertainties of adjusted parameters, as suggested by WD-code, are given in brackets.
\label{table5}}
%\tablewidth{25pt}
%\center
\tablehead{
\colhead{Parameter}        &  \colhead{V443~Cyg}      & \colhead{V456~Cyg}   & \colhead{V2107~Cyg} \\
\multicolumn{2}{l}{Adjusted:}
}
\startdata
$T_{\rm eff2}(K)$      &   6200(11)     & 6755(10)      & 15200(100)\\
$L_{1}/L_{1+2}(U)$     &   0.700(0.010) & 0.653(0.010)  & - \\
$L_{1}/L_{1+2}(B)$     &   0.694(0.007) & 0.693(0.007)  & 0.942(0.001)\\
$L_{1}/L_{1+2}(V)$     &   0.665(0.005) & 0.672(0.002)  & 0.938(0.001)\\
$L_{1}/L_{1+2}(R)$     &   0.658(0.004) & 0.639(0.005)  & 0.943(0.001)\\
$L_{1}/L_{1+2}(I)$     &        -       & 0.624(0.002)  &    -        \\
$L_{1}/L_{1+2}(Hp)$    &        -       &      -        & 0.938(0.002)\\
$\Omega_{\rm 1}$       &   6.78(0.04)   & 4.388(0.010)  & 4.17(0.01) \\
$\Omega_{\rm 2}$       &   8.18(0.04)   & 4.500(0.006)  & 7.32(0.01) \\
$\Omega_{\rm cr}$      &   3.73         & 3.49          & 2.99 \\
$r_{\rm 1}$(mean)      &   0.172(0.020) & 0.285(0.008)  & 0.280(0.003)\\
$r_{\rm 2}$(mean)      &   0.140(0.030) & 0.250(0.005)  & 0.092(0.005)\\
$i (^{o})$             &   89.9(0.5)    & 82.9(0.4)     & 86.2(0.5)\\
$e$                    &   0.0    & 0.0     & 0.045(0.03) \\
$w (^{o})$             &   -      & -       & 102 (2) \\
\hline
\multicolumn{2}{l}{Fixed:} \\
\hline
$P$ (days)             & 1.66220545    & 0.89119559  & 4.2845923\\
$T_0$                  & 2\,455\,454.6041  & 2\,455\,455.0958 & 2\,455\,259.3024 \\
$T_{\rm eff1}(K)$      &   6700        & 7750        & 22500 \\
$q (M_2/M_1)$          &   0.98       & 0.85       & 0.50       \\
$A_1=A_2$              & \multicolumn{3}{c}{1.0} \\
$g_1=g_2$              & \multicolumn{3}{c}{1.0} \\
$F_{\rm 1}=F_{\rm 2}$  & \multicolumn{3}{c}{1.0} \\
\enddata
\end{deluxetable}

\section{Discussion}

Table~6 lists the characteristics of the components and their orbits,  
as derived from the combined spectroscopic--photometric analysis. The  
proper motions were taken from Roeser, Demleitner, \& Schilbach (2010), 
while the distances and systemic velocities of the systems were derived in this study.
The space velocity components of the systems were calculated using the algorithm 
of Johnson \& Soderblom (1987) and corrected for the local standard of rest 
(U, V, W = 8.50, 13.38, 6.49) kms$^{-1}$ (Co{\c s}kuno{\v g}lu et al.,
2011). The total space velocities of V2107 Cyg, 
V443 Cyg and V456 Cyg are 63, 72 and 14 kms$^{-1}$, respectively.
Applying the N-body code of Dinescu et al.\ (1999) shows that the three 
systems move in the Galaxy in moderately eccentricity (Galactic) orbits ($e=0.17, 0.18,  
0.07$ for V2107~Cyg, V443~Cyg and V456~Cyg, respectively). These eccentricities show 
that the three systems belong to the thin-disc population. However, V2107~Cyg is at a distance of 
$1.5 \pm 0.5$~kpc to us, while V443~Cyg (at $0.6 \pm 0.2$~kpc) and 
V456~Cyg (at $0.50 \pm 0.03$~kpc) are much nearer to us. \footnote{Bolometric 
corrections from Strai\v{z}ys \& Kuriliene 1981, and $M_{V,\odot} = 4.75$~mag were used.}
The Mel'nik \& Efremov (1995) distance estimate of Cyg OB1 (1.4 kpc) 
and the Tetzlaff et al.\ (2010) space velocity components
($U, V, W = 45(2), -27(2), -7(2)$~kms$^{-1}$) are compatible with 
V2107~Cyg, that we consider a strong candidate member of this stellar 
group. The next subsection discusses shortly the implications for the 
star formation history of Cyg~OB1.

\begin{deluxetable}{ccccccc}
\setlength{\tabcolsep}{1pt}
\scriptsize{}
\tablecaption{Close binary stellar parameters of the program stars. Uncertainties
of parameters are given in brackets. \label{table6}}
\tablewidth{0pc}
\tablehead{ & \multicolumn{2}{c}{V443~Cyg} & \multicolumn{2}{c}{V456~Cyg}  & \multicolumn{2}{c}{V2107~Cyg} \\
Parameter  & Primary    & Secondary      & Primary & Secondary     & Primary & Secondary
}
\startdata
Sp      & F2/3V     & F8V      &  A2hA7mA4V  & F3V           &  B1III    & \nodata \\
\emph{M}\,(M$_\odot$)& 1.2(0.2)  & 1.1(0.3) &  1.86(0.06) & 1.58(0.05)    &  8.9(2.0) & 4.5(1.3) \\
\emph{R}\,(R$_\odot$)& 1.3(0.2)  & 1.1(0.2) &  1.68(0.02) & 1.47(0.02)    &  7.4(0.6) & 2.4(0.3) \\
\emph{a}\,(R$_\odot$)& \multicolumn{2}{c}{9.40(0.33)}& \multicolumn{2}{c}{5.88(0.06)} & \multicolumn{2}{c}{26.4(1.9)} \\
\emph{P}\,(days) & \multicolumn{2}{c}{1.66220545(43)}& \multicolumn{2}{c}{0.89119559(17)} & \multicolumn{2}{c}{4.2845923(25)} \\
\emph{i}\,($^{\circ}$) & \multicolumn{2}{c}{89.9(0.5)}  & \multicolumn{2}{c}{82.9(0.4)}  & \multicolumn{2}{c}{86.2(0.5)}     \\
\emph{q} $(M_2/M_1)$ & \multicolumn{2}{c}{0.98(0.09)} & \multicolumn{2}{c}{0.85(0.02)} & \multicolumn{2}{c}{0.50(0.06)} \\
$R_{L}$\,(R$_\odot$)    & \multicolumn{2}{c}{3.54} & \multicolumn{2}{c}{2.15} & \multicolumn{2}{c}{8.57} \\
\emph{e} & \multicolumn{2}{c}{0.0} & \multicolumn{2}{c}{0.0} & \multicolumn{2}{c}{$<$0.03} \\
$\log g$\,(cgs) & 4.25(0.23)& 4.42(0.26) & 4.258(0.027)& 4.302(0.027) & 3.65(0.17)& 4.3(0.2) \\
\emph{V}\,(mag) &  \multicolumn{2}{c}{12.31(0.03)} &  \multicolumn{2}{c}{10.80(0.03)} & \multicolumn{2}{c}{8.63(0.02)} \\
$B-V$\,(mag) & \multicolumn{2}{c}{0.51(0.09)} & \multicolumn{2}{c}{0.31(0.02)} & \multicolumn{2}{c}{0.10(0.05)} \\
$E(B-V)$\,(mag)& \multicolumn{2}{c}{0.13(0.15)} & \multicolumn{2}{c}{0.11(0.05)} & \multicolumn{2}{c}{0.335(0.045)} \\
$A_{\rm v}$\,(mag) &\multicolumn{2}{c}{0.42(0.15)}  &\multicolumn{2}{c}{0.35(0.08)} &\multicolumn{2}{c}{1.04(0.14)}  \\
$(B-V)_{\rm0}$\,(mag) &\multicolumn{2}{c}{0.38(0.06)} &\multicolumn{2}{c}{0.20(0.02)} &\multicolumn{2}{c}{-0.23(0.05)}  \\
$T_{\rm eff}$\,(K) & 6700(350) & 6200(550)  & 7750(100)  & 6755(400) & 22500(1500) & 15200(1600) \\
$\log$ \emph{L}\,(L$_\odot$)& 0.52(0.27) & 0.20(0.21) & 0.96(0.05) & 0.61(0.06) & 4.10(0.18) & 2.5(0.5) \\
$M_{\rm bol}$\,(mag) & 3.46(0.67) & 4.25(0.53) & 2.35(0.11) & 3.23(0.14) & --5.5(0.5) & --1.4(1.2) \\
$M_{\rm v}$\,(mag)  & 3.44(0.65) & 4.27(0.53) & 2.32(0.12) & 3.21(0.14) & --3.4(0.6) & --0.1(1.2) \\
\emph{BC}\,(mag)& 0.02(0.02) & --0.02(0.05)& 0.03(0.01)& 0.02(0.01) & --2.2(0.2) & --1.3(0.2) \\
$K_{\rm 1,2}$\,(km\,s$^{-1}$)& 117.4(6.2) & 120.2(4.2)  & 152.1(1.7) & 179.4(1.7) & 104(1.5) & 207(20) \\
$V_{\gamma}$\,(km\,s$^{-1}$) & \multicolumn{2}{c}{-19.7(0.7)} & \multicolumn{2}{c}{--2.7(1.0)} & \multicolumn{2}{c}{--6.3(0.2)}  \\
V$_{synch}$\,(km\,s$^{-1}$)   & 41(7) & 33(6) &  95(1)  & 83(1) & 87(7) & 29(4)  \\
V$_{\rm rot}$\,(km\,s$^{-1}$) & \nodata     & \nodata     & 100(10) & \nodata     & 84(4) & \nodata    \\
\emph{d}\,(pc) & \multicolumn{2}{c}{600(200)} & \multicolumn{2}{c}{500(30)} & \multicolumn{2}{c}{1500(500)}\\
$\mu_\alpha cos\delta$, $\mu_\delta$\,(mas yr$^{-1}$) & \multicolumn{2}{c}{--11.1(2.2), --15.8(2.2)} & \multicolumn{2}{c}{5.6(1.2),
1.3(1.1)}  & \multicolumn{2}{c}{--4.8(1.3), --6.5(1.3)}  \\
$U$\,(km\,s$^{-1}$) & \multicolumn{2}{c}{48.98(22.39)} & \multicolumn{2}{c}{--10.66(2.73)} & \multicolumn{2}{c}{49.76(19.06)} \\
$V$\,(km\,s$^{-1}$) & \multicolumn{2}{c}{--31.05(5.02)} & \multicolumn{2}{c}{--0.46(1.14)} & \multicolumn{2}{c}{--20.79(5.46)} \\
$W$\,(km\,s$^{-1}$) & \multicolumn{2}{c}{--0.57(6.22)} & \multicolumn{2}{c}{--9.02(2.84)} & \multicolumn{2}{c}{3.14(8.70)} \\
\enddata

\end{deluxetable}

\subsection{V2107~Cyg and the age of Cyg~OB1}

The primary component is obviously near the end of the main-sequence, 
while the secondary is near the zero-age main-sequence. The fundamental stellar parameters 
of the secondary suggest a spectral type B5 V. In view of the light contribution of about 
six per cent to the total light, core of the diluted Hydrogen lines would be three per cent 
deep in the observed spectra and the strongest metal lines, Mg II\,$\lambda$4481 would be 
one per cent deep in case of rotation synchronized with the orbit. In order to detect these lines
with confidence one requires on the order of 15 spectra with S/N of 100.
Fig.~9 shows the location of the components in the $\log T_{eff}$ -- $\log L$ plane, and 
isochrones calculated using the web 
interface\footnote{http://stev.oapd.inaf.it/YZVAR/cgi-bin/form}. 
A solar mixture was assumed, i.e. Z=0.017, Y=0.26. Despite the present    
uncertainties, the age of the system can be constrained to $20 \pm 5$~Myr  
due to the favourable position of the primary component. The age estimate  
of 7.5 Myr for the group as a whole (Mel'nik \& Efremov 1995) suggests that 
star formation may have been on-going for at least $10^ 7$~yr. 
Hence, as well in Cyg~OB1 as in Cyg~OB2 (Comeron \& Pasquali 2012), there 
is evidence in favour of sustained star formation for well over 10 Myr. 
 
This study identifies V2107\,Cyg as an extremely interesting binary star 
with the potential to constrain the duration of the star formation process 
in the Cyg OB1 association with precision. More extensive high-resolution 
spectroscopy, as already obtained at the mid-eclipses, is required to 
constrain $K_2$ precisely (and thus the mass of the primary). Moreover,  
with the secondary component of V2107\,Cyg occulted for six consecutive 
hours, the intrinsic line profile variations of the primary component can 
be studied even without the (weak) contamination and dilution by the 
secondary star. It has been noticed, in secondary eclipse, that many metal   
lines show a peculiar line core with a similar asymmetry, as illustrated
for the Si~{\sc iii} triplet in Fig.~5. Spectroscopic monitoring should 
confirm that this is due to non-radial pulsations.

\begin{figure}
\begin{center}
\epsscale{0.5}
\plotone{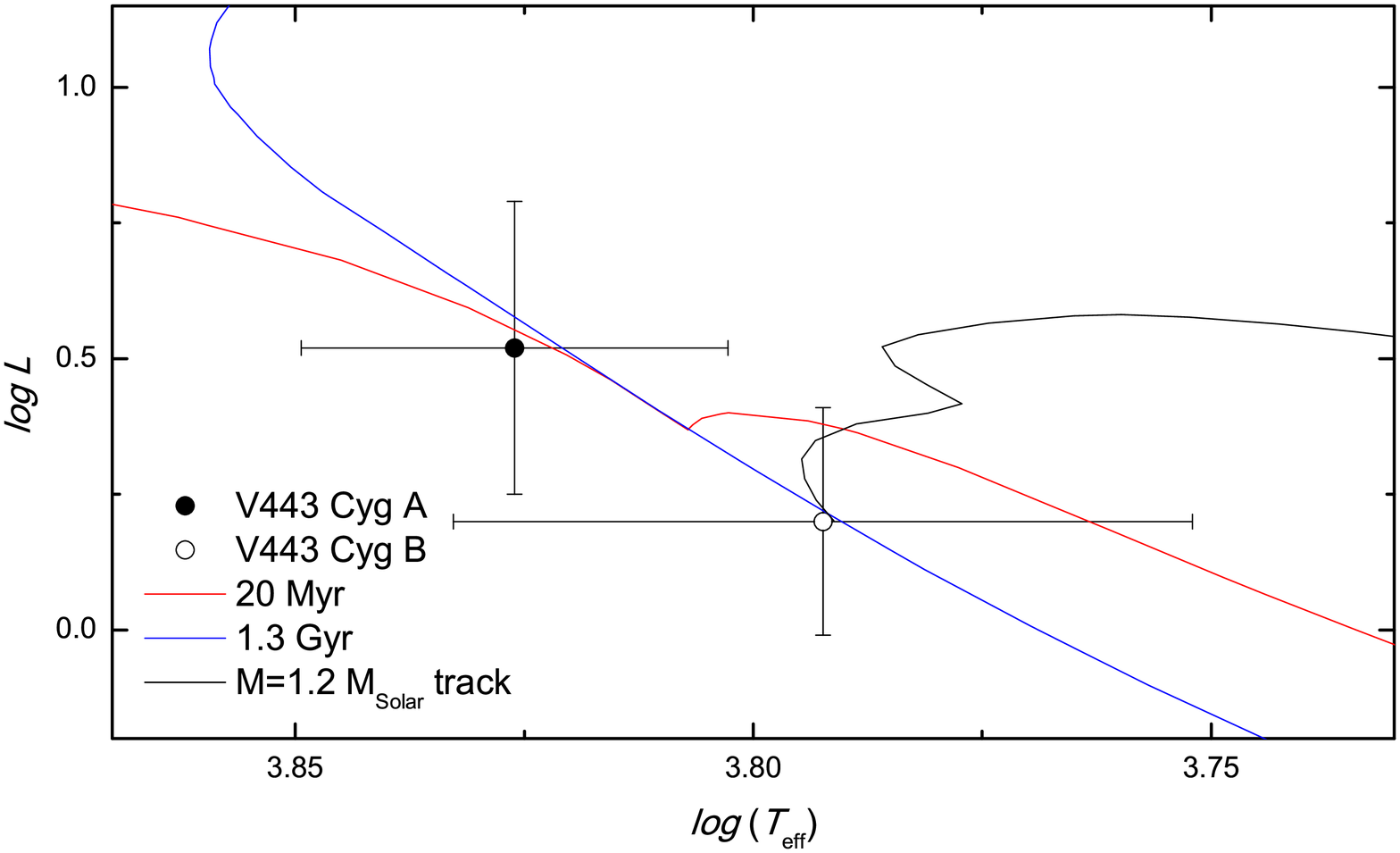} \\
%\vspace{5mm}
\plotone{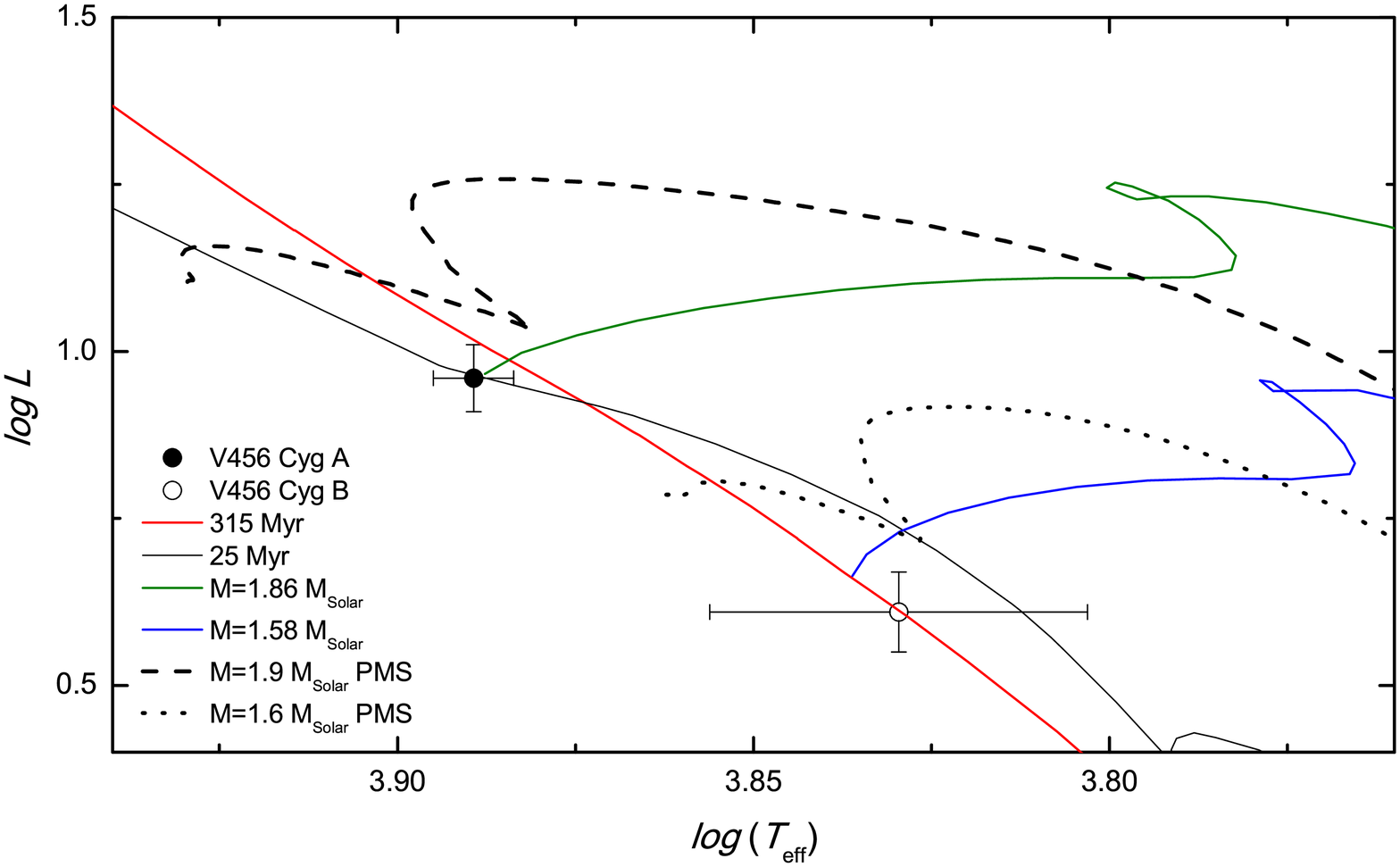} \\
%\vspace{5mm}
\plotone{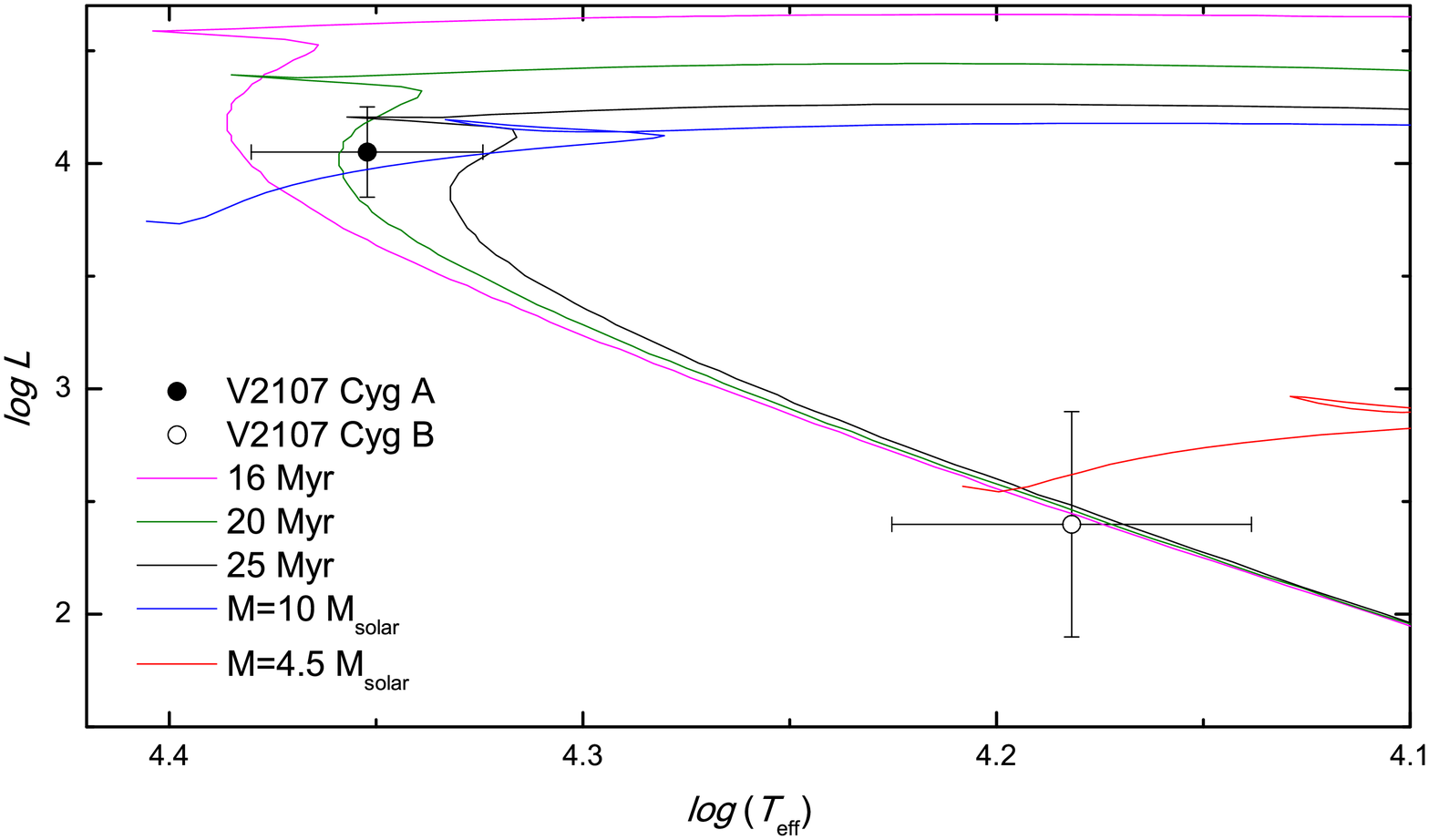} \\
\caption{Locations of program stars in the $\log~L$ -- $\log~T_{eff}$ plane together 
with evolutionary tracks and isochrone curves. Tracks and isochrones for V456~Cyg 
are calculated for Z=0.035 (full line) and Z=0.020 (dashed, PMS), respectively.}\label{fig9}
\end{center}
\end{figure}

\clearpage

\acknowledgments

This work is fully supported by the Scientific and Technological Research
Council (T\"{U}B\.{I}TAK)
of Turkey under the project code 109T449. The spectroscopic observations are
granted by the T\"{U}B\.{I}TAK National Observatory with the project code
10BRTT150-30-0. The photometric observations are granted by Ulupinar Observatory
of \c{C}anakkale Onsekiz Mart University. Part of the spectroscopic observations
of V2107~Cyg is based on observations obtained with the HERMES spectrograph,
which is supported by the Fund for Scientific Research of Flanders (FWO),
Belgium , the Research Council of K.U.Leuven, Belgium, the Fonds National
Recherches Scientific (FNRS), Belgium, the Royal Observatory of Belgium, the
Observatoire de Gen\`{e}ve, Switzerland and the Thüringer Landessternwarte
Tautenburg, Germany. We thank the observers Ben Devries, 
Yves Fr\'{e}mat, Marie Hrudkova and Peter Papics. This research made use of {\sc
simbad} and {\sc aladin}. We made use of data from the UVES Paranal Observatory
project (ESO DDT Program ID 266.D-5655).


\begin{thebibliography}{}
\bibitem[\protect\citeauthoryear{Bagnulo}{2003}]{B03} Bagnulo S., Jehin E., Ledoux C., Cabanac R., Melo C., Gilmozzi R. et al., 2003, ESO Messenger, 114, 10
\bibitem[\protect\citeauthoryear{Bertelli}{2009}]{B04} Bertelli G., Nasi E., Girardi L., Marigo P., 2009, A\&A, 508, 355
\bibitem[\protect\citeauthoryear{Bikmaev}{2005}]{B05} Bikmaev I., Sakhibullin N., Musaev F., Aslan Z., 2005, http://www.tug.tubitak.gov.tr
\bibitem[\protect\citeauthoryear{Bissell}{1975}]{B06} Bissell A.\,F., Ferguson R.\,A., 1975, The Statistician, 24, 79
\bibitem[\protect\citeauthoryear{Boeche}{2004}]{B07} Boeche C., Munari U., Tomasella L., Barbon R., 2004, A\&A, 415, 145
\bibitem[\protect\citeauthoryear{Cenarro}{2007}]{C01} Cenarro A.\,J., Peletier R.\,F., Sanchez-Blazquez P., et al., 2007, MNRAS, 374, 664
\bibitem[\protect\citeauthoryear{Comeron}{2012}]{C02} Comeron F., Pasquali A., 2012, A\&A, 543,A101
\bibitem[\protect\citeauthoryear{Co{\c s}kuno{\v g}lu et al.}{2011}]{C03} Co{\c s}kuno{\v g}lu B., Ak S., Bilir S., Karaali S., et al., 2011, MNRAS, 412, 1237
\bibitem[\protect\citeauthoryear{Cunha}{2000}]{C04} Cunha K., Smith V.\,V., Boesgaard A.\,M., Lambert D.\,L., 2000, ApJ, 530, 939
\bibitem[\protect\citeauthoryear{Dinescu, Girard, van Altena}{1999}]{D01} Dinescu D.\,I., Girard T.\,M., van Altena W.\,F., 1999, AJ, 117, 1792
\bibitem[\protect\citeauthoryear{Edvardsson}{1993}]{E01} Edvardsson B., Andersen J., Gustafsson B., Lambert D.\,L., Nissen P.\,E., Tomkin J., 1993, A\&A, 275, 101
\bibitem[\protect\citeauthoryear{ESA}{1997}]{E02} ESA, 1997, The Hipparcos and Tycho Catalogues, ESA SP-1200
\bibitem[\protect\citeauthoryear{Gratton}{1996}]{G01} Gratton R.\,G., Carretta E., Castelli F., 1996, A\&A, 314, 191
\bibitem[\protect\citeauthoryear{Gray}{1992}]{G02} Gray D.\,F., 1992, The Observation and Analysis of Stellar Photospheres (2nd ed.; Cambridge: Cambridge Univ. Press) 
\bibitem[\protect\citeauthoryear{Gray}{2006}]{G03} Gray R.\,O., Corbally C.\,F., Garrison R.\,F., Mc Fadden M.\,T., O'Donaghue A.\,A., Bubar E.\,J., 2006, A\&AS, 209, 8902
\bibitem[\protect\citeauthoryear{Gray}{1988}]{G04} Gray R.\,O., 1988, AJ, 95, 220
\bibitem[\protect\citeauthoryear{Hadrava}{1995}]{H01} Hadrava P., 1995, A\&AS, 114, 393
\bibitem[\protect\citeauthoryear{Hoffman}{2008}]{H02} Hoffman D. I., Harrison T. E., Coughlin J. L., McNamara B. J., Holtzman J. A., Taylor G. E., Vestrand W. T., 2008, AJ, 136, 1067
\bibitem[\protect\citeauthoryear{Ilijic}{2003}]{I03} Iliji\'{c} S., 2003, Master thesis, University of Zagreb
\bibitem[\protect\citeauthoryear{Kiminki}{2012}]{K01} Kiminki Daniel C., Kobulnicky Henry A., 2012, ApJ, 751, 4
\bibitem[\protect\citeauthoryear{Kobulnicky}{2012}]{K02} Kobulnicky Henry A., Smullen Rachel A., Kiminki Daniel C., Runnoe Jessie C., Wood Earl S., Long Garrett, Alexander Michael J., Lundquist Michael J., Vargas-Alvarez Carlos, 2012, ApJ, 756, 50
\bibitem[\protect\citeauthoryear{Kurucz}{1993}]{K03} Kurucz R.\,L., 1993, in ASP Conf.Ser.44, Peculiar versus Normal Phenomena in A-type and Related Stars, ed. M.M.Dworetsky, F.Castelli, R. Faraggiana (San Francisco, CA: ASP), 87
\bibitem[\protect\citeauthoryear{Johnson}{1987}]{J01} Johnson D.\,R.\,H., Soderblom D.\,R., 1987, AJ, 93, 864
\bibitem[\protect\citeauthoryear{Lucy}{1967}]{L02} Lucy L.\,B., 1967, Z. Astrophys., 65, 89
\bibitem[\protect\citeauthoryear{Mahy}{2013}]{M01} Mahy L., Rauw G., De Becker M., Eenens P., Flores C. A., 2013, A\&A, 550, 27
\bibitem[\protect\citeauthoryear{Melnik}{1995}]{M02} Mel'nik A.\,M., Efremov Yu.\,N., 1995, AstL, 21, 10
\bibitem[\protect\citeauthoryear{Mercier}{1957}]{M03} Mercier J.P., 1957, Journal des Observateurs, 40, 12
\bibitem[\protect\citeauthoryear{Morgan}{1955}]{M04} Morgan W.\,W., Code A.\,D., Whitford A.\,E., 1955, ApJS, 2, 41
\bibitem[\protect\citeauthoryear{Nelson}{2011}]{N01} Nelson R.\,H., 2011, Inf. Bull. on Var. Stars, 5994
\bibitem[\protect\citeauthoryear{Nieva}{2007}]{N02} Nieva M.\,F., Przybilla N., 2007, A\&A 467, 295
\bibitem[\protect\citeauthoryear{Prsa}{2005}]{P02} Pr\v{s}a A., Zwitter T., 2005, ApJ, 628, 426
\bibitem[\protect\citeauthoryear{Raskin}{2011}]{R01} Raskin G., van Winckel H., Hensberge H., et al., 2011, A\&A, 526, 69
\bibitem[\protect\citeauthoryear{Reipurth}{2008}]{R02} Reipurth B., Schneider N., 2008, Handbook of Star Forming Regions, Volume I: The Northern Sky ASP Monograph Publications, Vol. 4. Edited by Bo Reipurth, p.36
\bibitem[\protect\citeauthoryear{Roeser}{2010}]{R04} Roeser S., Demleitner M., Schilbach E., 2010, AJ, 139, 2440
\bibitem[\protect\citeauthoryear{Roman}{1951}]{R05} Roman N. G., 1951, ApJ, 114, 492
\bibitem[\protect\citeauthoryear{Rucinski}{1973}]{R08} Rucinski S.M., 1973, Acta Astron., 23, 79
\bibitem[\protect\citeauthoryear{Straizys}{1981}]{S03} Strai\v{z}ys V., Kuriliene G., 1981, Ap\&SS, 80, 353
\bibitem[\protect\citeauthoryear{Tetzlaff}{2010}]{T01} Tetzlaff N., Neuh\"{a}user R., Hohle M. M., Maciejewski G., 2010, MNRAS, 402, 2369
\bibitem[\protect\citeauthoryear{vanHamme}{1993}]{V01} van Hamme W., 1993, AJ, 106, 2096
\bibitem[\protect\citeauthoryear{vrancken}{1997}]{V02} Vrancken M., Hensberge H., David M., Verschueren W., 1997, A\&A, 320, 878
\bibitem[\protect\citeauthoryear{Walborn}{1971}]{W01} Walborn N.\,R., 1971, ApJS, 23, 257
\bibitem[\protect\citeauthoryear{Wilson}{1971}]{W02} Wilson R.\,E., Devinney E.\,J., 1971, ApJ, 166, 605
\bibitem[\protect\citeauthoryear{Wilson}{1990}]{W03} Wilson R.\,E., 1990, ApJ, 356, 613
\bibitem[\protect\citeauthoryear{Wozniak}{2004}]{W04} Wozniak P.\,R., Vestrand W.\,T., Akerlof C.\,W., et al., 2004, AJ, 127, 2436
\bibitem[\protect\citeauthoryear{Zakirov}{1999}]{Z01} Zakirov M.\,M., 1999, AstL, 25, 229
\bibitem[\protect\citeauthoryear{Zakirov}{2005}]{Z02} Zakirov M.\,M., Eshankulova M.\,U., 2005, KFTN, 21, 441
\bibitem[\protect\citeauthoryear{Zakirov}{2006}]{Z03} Zakirov M.\,M., Eshankulova M.\,U., 2006, KFTN, 22, 363
\end{thebibliography}
\end{document}